\documentclass{article} % For LaTeX2e
\usepackage{nips15submit_e,times}
\usepackage{hyperref}
\usepackage{url}
\usepackage{amsmath,amssymb}
\usepackage{cite}
\usepackage{graphicx}
\title{Approximate Inference for Time-varying Interactions and Macroscopic Dynamics of Neural Populations}

\author{ Christian Donner \\
Bernstein Center for Computation Neuroscience Berlin\\
Neural Information Processing Group \\
Technische Universit\"at Berlin \\
\texttt{christian.donner@bccn-berlin.de} \\
\And
Klaus Obermayer \\
Bernstein Center for Computation Neuroscience Berlin\\
Neural Information Processing Group \\
Technische Universit\"at Berlin \\
\texttt{oby@ni.tu-berlin.de}
\AND
Hideaki Shimazaki \\
RIKEN Brain Science Institute \\
\texttt{shimazaki@brain.riken.jp} \\
}

\nipsfinalcopy % Uncomment for camera-ready version

\begin{document}

\maketitle

\begin{abstract}
The models in statistical physics such as an Ising model offer a convenient way to characterize stationary activity of neural populations. Such stationary activity of neurons may be expected for recordings from \textit{in vitro} slices or anesthetized animals. However, modeling activity of cortical circuitries of awake animals has been more challenging because both spike-rates and interactions can change according to sensory stimulation, behavior, or an internal state of the brain. Previous approaches modeling the dynamics of neural interactions suffer from computational cost; therefore, its application was limited to only a dozen neurons. Here by introducing multiple analytic approximation methods to a state-space model of neural population activity, we make it possible to estimate dynamic pairwise interactions of up to $60$ neurons. More specifically, we applied the pseudolikelihood approximation to the state-space model, and combined it with the Bethe or TAP mean-field approximation to make the sequential Bayesian estimation of the model parameters possible. The large-scale analysis allows us to investigate dynamics of macroscopic properties of neural circuitries underlying stimulus processing and behavior. We show that the model accurately estimates dynamics of network properties such as sparseness, entropy, and heat capacity by simulated data, and demonstrate utilities of these measures by analyzing activity of monkey V4 neurons as well as a simulated balanced network of spiking neurons.
\end{abstract}

% Use "Eq" instead of "Equation" for equation citations.
\section*{Introduction}
Activity patterns of neuronal populations are constrained by biological mechanisms such as biophysical properties of each neuron (e.g., synaptic integration and spike generation \cite{london2005dendritic,de2007correlation}) and their anatomical connections \cite{reyes2003synchrony}. The characteristic correlations among neurons imposed by the biological mechanisms interplay with statistics of sensory inputs, and influence how the sensory information is represented in the population activity \cite{pitkow2012decorrelation,kenet2003spontaneously,luczak2009spontaneous}. Thus accurate assessment of the neural correlations in ongoing and evoked activities is a key to understand the underlying biological mechanisms and their coding principles. 
\par
The number of possible activity patterns increases combinatorially with the number of neurons analyzed. The maximum entropy (ME) principle and derived ME models - known as the pairwise ME model or the Ising model - have been used to explain neural population activities using fewer activity features such as event rates or correlations between pairs of neurons \cite{shlens2006structure, schneidman2006weak}. This approach has been employed to explain not only the activity of neuronal networks but also other types of biological networks \cite{lezon2006using,mora2010maximum,bialek2012statistical}. For large networks, however, exact inference of these models becomes computationally infeasible. Thus researchers have employed approximation methods \cite{sessak2009small,roudi2009statistical,roudi2009ising,sohl2011new,schaub2012ising,cocco2012adaptive,haslinger2013missing}. While they successfully extended the number of neurons that could be analyzed, it was pointed out that the pairwise ME model might fail to explain large neural populations because the effect of higher-order interactions may become prominent \cite{roudi2009pairwise,ganmor2011sparse,tkavcik2014searching}. Another fundamental problem of the conventional ME models is that these models assume temporarily constant spike rates for individual neurons. The assumption of stationary spike-rates is invalid, e.g., when \textit{in vivo} activity is recorded while an animal performs a behavioral task. Ignoring such dynamics might result in erroneous model estimates and misleading interpretations on their correlations \cite{brody1999correlations,grun2009data,renart2010asynchronous,roudi2011mean,tyrcha2013effect}. Moreover neural correlations themselves likely organize dynamically during behavior and cognition, which can be independent from changes in the spike rates of individual neurons \cite{vaadia1995dynamics,riehle1997spike,sakurai2006dynamic}. The time-dependence of neural activity may be explained by including stimulus signals in the model, e.g., for analyses of early sensory cells \cite{granot2013stimulus}. However, the approach may become impractical when analyzing neurons in higher brain areas in which receptive fields of neurons are not easily characterized. Thus it remains to be examined how much the pairwise ME model can explain the data if the inappropriate stationary assumption is removed.

\par
The state-space analysis \cite{chen2013state} offers a general framework to model time-series data as observations driven by an unobserved latent state process. The underlying state changes are uncovered by a sequential estimation method from the noisy measurements. While observations of neuronal activity are often characterized by point events (spikes), a series of studies have established the nonlinear recursive Bayesian estimation of the underlying state that drives the event activity \cite{brown1998statistical,smith2003estimating,eden2004dynamic}. The method successfully estimated an animal's position from population activity of hippocampal place cells \cite{brown1998statistical}, or estimate arm trajectories from neurons in the monkey motor cortex \cite{truccolo2005point,srinivasan2006state}. Recently, this framework has been extended to the analysis of population activity \cite{shimazaki2009state,shimazaki2012state,shimazaki2013single}. In addition to the point estimates of interaction parameters suggested by earlier studies \cite{kolar2010estimating,long2011statistical,kass2011assessment}, the state-space analysis provides credible intervals of those estimates through the recursive Bayesian fitting algorithm. 
\par
Nevertheless, as previously mentioned, the state-space model of a neural population was restricted by its computational cost. Therefore, it could be utilized to analyze only small populations ($N \leq 15$). Recent advances in electrophysiological and optical recording techniques from a large number of neurons \textit{in vivo} under free moving or virtual reality settings challenge these analysis methods. Thus the challenge is to make it possible to fit the exponentially complex state-space model to such large-scale data. For this goal, we need to incorporate approximation methods into the sequential Bayesian algorithm. More specifically, we need good approximations of mean and variance of the model parameters required in the approximate Bayesian scheme. These approximation methods must be analytical to avoid impractical computation time. By doing so we will be able to directly estimate all time-varying interactions of a large neural population. Such a model will serve as benchmark for alternative unsupervised methods that aim to capture low-dimensional, time-dependent latent structure of the pairwise interactions \cite{hayashi2009dynamic,effenberger2015discovery,hirayama2016sparse} (see also \cite{byron2009gaussian,cunningham2014dimensionality,okun2015diverse} for other dimension reduction methods for neuroscience data).

\par
Here by combining the state-space model proposed in  \cite{shimazaki2009state,shimazaki2012state,shimazaki2013single} with analytic approximation methods, we provide a framework for estimating interactions of neuronal populations consisting of up to $60$ neurons. To find the mean we used the pseudolikelihood approximation method. To approximate the variance, we provide two alternative methods: the Bethe or the mean-field approximation. The Bayesian analysis methods for larger networks of neurons allow us to better understand macroscopic states of a neural population, such as entropy, free energy and sensitivity, all in a time-resolved manner and with credible intervals. Thus the model provides a new way to investigate effects of stimuli and behavior on activity of neuronal populations. It is expected to provide observations that give us insights into the underlying circuitry and its computation.

\section*{Materials and Methods}
To clarify the problem of large-scale analysis on dynamic population activity, we first formulate the state-space model and its estimation method originally investigated in \cite{shimazaki2009state,shimazaki2012state} in the next subsection. Then we describe how to introduce approximation methods to the state-space model in order to overcome the limitation of the model and make the large-scale analysis possible. The custom-made Python programs are provided on GitHub (\href{https://github.com/christiando/ssll_lib}{https://github.com/christiando/ssll\_lib}).

\subsection*{The state-space analysis of neural population activity}
\textbf{Spike data} To investigate how neuronal activities realize perception, cognition, and behavior, neurophysiologists record timing of neuronal spiking activity over the course of a behavioral paradigm designed to test specific hypotheses. Typically, these experiments are repeated multiple times under the same experimental conditions to uncover common neuronal dynamics related to the behavioral paradigm from stochastic spiking activities. We assume that neural data is composed of repeated measurements ($R$ times) of spike timing recorded from $N$ neurons simultaneously. Hereafter repetition is termed trial. To analyze activity patterns of neurons, we discretize the parallel spike sequences into $T$ time bins with bin size $\Delta$, and represent the population activity by a set of binary variables. For neurons $n=1,\ldots,N$, time bins $t=1,\ldots,T$, and trials $r=1,\ldots,R$, the neural activity is represented by a binary variable $X_n^{r,t}$, where $X_n^{r,t}=1$ when neuron $n$ spiked in time bin $t$ and trial $r$; and $X_n^{r,t} = 0$ otherwise. Hence, we describe the whole data as a $N\times R \times T$ dimensional binary matrix. The activity pattern of $N$ neurons at time bin $t$ and trial $r$ is a vector, $\mathbf{X}^{r,t}=(X_1^{r,t},\ldots,X_N^{r,t})^\prime$.  Similarly, $\mathbf{X}^t=(\mathbf{X}^{1,t},\ldots,\mathbf{X}^{R,t})$ summarizes observations for all neurons $1,\ldots,N$ and all trials $1,\ldots,R$ at time bin $t$. Finally, $\mathbf{X}^{t_1:t_2}=(\mathbf{X}^{t_1},\ldots,\mathbf{X}^{t_2})$ denotes the observations from time bin $t_1$ to $t_2$.

\bigskip \noindent 
\textbf{State-space model of neural population activity} We assume a state-space model of dynamic population activity composed of two submodels; an observation model and a state model. First, the observation model specifies the probability distribution of population activity patterns using state variables, whereas the latter dictates how those state variables change. Here we construct the observation model using the exponential family distribution considering up to pairwise interactions of neurons' activities, 
\begin{equation}\label{eq:observationmodel}
p(\mathbf{x}\vert \boldsymbol{\theta}_t) 
 = \exp\left[ \sum_{i=1}^N\theta^t_i x_i + \sum_{j> i}\theta^t_{ij}x_jx_i - \psi_t (\boldsymbol\theta_t) \right],
\end{equation}
where $\psi_t(\boldsymbol\theta_t)$ is a log normalization term (a.k.a. log partition function). The model contains $d = N + N(N-1)/2$ parameters $\lbrace  \theta_i^t \rbrace, \lbrace \theta_{ij}^t \rbrace$ known as natural or canonical parameters of an exponential family distribution. In statistical mechanics, this model is named ``Ising model'', where the vector $\mathbf{x}$ represents a spin configuration (up or down). There, the natural parameters $\lbrace \theta_i^t \rbrace, \lbrace\theta_{ij}^t \rbrace$ represent external magnetic field and interactions among the spins, and may be denoted as $\lbrace h_{i}\rbrace,\lbrace J_{ij}\rbrace$ conventionally. Here we consider these parameters to be time-dependent, and refer to them as state variables of the state-space model. By introducing the $d$-dimensional state vector $\boldsymbol\theta_t = (\theta_1^t,\ldots,\theta_N^t,\theta^t_{1,2},\ldots,\theta^t_{N-1,N})^\prime$, and the feature vector $\mathbf{F}(\mathbf{x})=(x_1,\ldots,x_N,x_1x_2,\ldots,x_{N-1}x_{N})^\prime$, the model of Eq~\ref{eq:observationmodel} is written concisely as
$p(\mathbf{x}\vert \boldsymbol{\theta}_t) = \exp [ \boldsymbol\theta_t^\prime\mathbf{F}(\mathbf{x}) - \psi_t(\boldsymbol\theta_t) ]$. The resulting log partition function is then given by  
\begin{align} \label{eq:freeenergy}
\psi_t(\boldsymbol\theta_t) = \log\sum_\mathbf{x}\exp[\boldsymbol\theta_t^\prime \, \mathbf{F}(\mathbf{x})].
\end{align}
In statistical mechanics, $\psi_t$ is known as the free energy. Note that it specifies the probability that all neurons are simultaneously silent because $p(\mathbf{0}\vert \boldsymbol{\theta}_t )=\exp[-\psi_t(\boldsymbol\theta_t)]$. This model considers individual and pairwise activity of neurons. Hence, we will refer to it as the \emph{pairwise observation model} in the following.

\par
Next, the state model considers that dynamics of the latent state $\boldsymbol\theta_t$ is described by a random walk
\begin{align}\label{eq:statemodel}
\boldsymbol\theta_t = \boldsymbol\theta_{t-1} + \boldsymbol{\xi}_t(\lambda),
\end{align}
where $\boldsymbol{\xi}_t$ is a random vector drawn from a multivariate normal distribution $\mathcal{N}(\boldsymbol{0},\mathbf{Q})$, and $\mathbf{Q}$ is a diagonal covariance matrix. Here we assume that entries of the diagonal of the inverse matrix $\mathbf{Q}^{-1}$ are given by a scalar $\lambda$ that determines precision of the noise for all elements. For the initial time bin we set the density to $p(\boldsymbol\theta_{1}) = \mathcal{N}(\boldsymbol\mu,\boldsymbol\Sigma)$.

\par
It should be noted that here we model the neural dynamics as a \textit{quasistatic} process, similarly to the classical analysis on dynamics of a thermodynamic system, e.g., a heat engine (see also \cite{shimazaki2015neurons}): At each time $t$, we presume that neural activity is sampled from the \textit{equilibrium} distribution (Eq~\ref{eq:observationmodel}), which is the same across the trials (across-trial stationarity). The free energy (Eq~\ref{eq:freeenergy}) is also defined in the same manner as in the classical thermodynamics. We emphasize that the quasistatic process is a simplified view of the neural dynamics. See Discussion for possible extensions of the model. 

\bigskip \noindent 
\textbf{Estimating the state-space model} Given the data $\mathbf{X}^{1:T}$, our goal is to jointly estimate the posterior density of the latent states and the optimal noise precision $\lambda$. By denoting hyperparameters of the model as $\mathbf{w}=(\lambda,\boldsymbol\mu,\boldsymbol\Sigma)$, the posterior density of the state process writes as 
\begin{equation}\label{eq:smoother}
p(\boldsymbol{\theta}_{1:T}|\mathbf{X}^{1:T},\mathbf{w})
= \frac{p(\mathbf{X}^{1:T} | \boldsymbol{\theta}_{1:T}) p(\boldsymbol{\theta}_{1:T} ; \mathbf{w}) } {p(\mathbf{X}^{1:T}; \mathbf{w})},
\end{equation}
where the first component in the numerator is constructed from the observation model, and the second component from the state model. In the next section, we provide the iterative method to construct this posterior density by approximating it by a Gaussian distribution (the Laplace approximation). 
The posterior density depends on the choice of the parameters $\mathbf{w}$. The optimal $\mathbf{w}$ maximizes the marginal likelihood, a.k.a. evidence, that appears in the denominator in Eq~\ref{eq:smoother}, given by
\begin{equation}\label{eq:marglikelihood}
l(\mathbf{X}^{1:T} |\mathbf{w})= p(\mathbf{X}^{1}|\boldsymbol{\mu}, \boldsymbol{\Sigma})
\prod_{t=2}^T p(\mathbf{X}^{t}|\mathbf{X}^{1:t-1}, \lambda).
\end{equation}
This approach is called the empirical Bayes method. In this study, we optimize noise precision $\lambda$ and mean $\boldsymbol\mu$ of the initial distribution as described below while values for the covariance $\boldsymbol\Sigma$ are fixed. For fitting in the subsequent analyses, we set initial values as $\lambda=100$ and $\mathbf{\Sigma}=10\mathbf{I}$. For initial value of $\boldsymbol\mu$ we computed the vector $\boldsymbol\theta$ from time and trial averaged data, assuming $\lbrace\theta_{ij}^t\rbrace=\boldsymbol{0}$.

The optimization is achieved by an EM-algorithm combined with recursive Bayesian filtering/smoothing algorithms  \cite{shumway1982approach,smith2003estimating}. In this approach, we alternately perform construction of the posterior density (Eq~\ref{eq:smoother}, E-step) and optimization of the hyperparameters (M-step) until the marginal likelihood (Eq~\ref{eq:marglikelihood}) saturates. In order to update the hyperparameters to new values $\mathbf{\mathbf{w}^{\ast}}$ from old values $\mathbf{\mathbf{w}}$ in the M-step, a lower bound of the marginal likelihood is maximized. This lower bound is obtained by applying the Jensen’s inequality to the marginal likelihood:
\begin{align}
l(\mathbf{X}^{1:T} |\mathbf{w}^{\ast}) 
& = \log \int p(\mathbf{X}^{1:T},\boldsymbol{\theta}_{1:T}| \mathbf{w}^{\ast}) d\boldsymbol{\theta}_{1:T} \nonumber\\
& =
\log \left\langle
\frac{p(\mathbf{X}^{1:T},\boldsymbol{\theta}_{1:T}| \mathbf{w}^{\ast})}
{p(\boldsymbol\theta_{1:T}|\mathbf{X}^{1:T},\mathbf{w})}
\right\rangle_{\boldsymbol\theta_{1:T}|\mathbf{X}^{1:T},\mathbf{w}} \nonumber\\
& \geq 
\left\langle
\log \frac{p(\mathbf{X}^{1:T},\boldsymbol{\theta}_{1:T}| \mathbf{w}^{\ast})}
{p(\boldsymbol\theta_{1:T}|\mathbf{X}^{1:T},\mathbf{w})}
\right\rangle_{\boldsymbol\theta_{1:T}|\mathbf{X}^{1:T},\mathbf{w}} \nonumber\\
& = 
\left\langle
\log p(\mathbf{X}^{1:T},\boldsymbol{\theta}_{1:T}| \mathbf{w}^{\ast})
\right\rangle_{\boldsymbol\theta_{1:T}|\mathbf{X}^{1:T},\mathbf{w}}
-\left\langle 
\log p(\boldsymbol\theta_{1:T}|\mathbf{X}^{1:T},\mathbf{w})
\right\rangle_{\boldsymbol\theta_{1:T}|\mathbf{X}^{1:T},\mathbf{w}}
\end{align}
Here $\left\langle\centerdot\right\rangle_{\boldsymbol\theta_{1:T}|\mathbf{X}^{1:T},\mathbf{w}}$ is expectation by the posterior density of the state variables (Eq~\ref{eq:smoother}). 
In order to maximize the lower bound w.r.t. the new hyperparameters $\mathbf{w}^{\ast}$, we only need to maximize the first term, $q(\mathbf{w}^{\ast}|\mathbf{w}) \equiv \left\langle \log  p\left(\mathbf{X}^{1:T},\boldsymbol{\mathbf{\theta}}_{1:T}|\mathbf{\mathbf{w}^{\ast}}\right)\right\rangle_{\boldsymbol\theta_{1:T}|\mathbf{X}^{1:T},\mathbf{w}} $. This term is called expected complete data log-likelihood, where the expectation is taken by the posterior density with the old $\mathbf{\mathbf{w}}$. It is computed as
\begin{align} \label{eq:Q-function}
q(\mathbf{w}^{\ast}|\mathbf{w}) =& \sum_{t=1}^{T} \sum_{r=1}^{R} 
\langle
\boldsymbol{\theta}_t^\prime \mathbf{F}(\mathbf{X}^{t,r}) - \psi(\boldsymbol{\theta}_t) 
\rangle_{\boldsymbol\theta_{1:T}|\mathbf{X}^{1:T},\mathbf{w}} \nonumber\\
& -\frac{1}{2}\log{|2\pi \boldsymbol{\Sigma}^{\ast}|} - \frac{1}{2}\left\langle\left(\boldsymbol{\theta}_{1} 
-\boldsymbol{\mu}^{\ast}\right)^{\prime}\boldsymbol{\Sigma}^{\ast-1}\left(\boldsymbol{\theta}_{1}-\boldsymbol{\mu}^{\ast}\right)\right\rangle_{\boldsymbol\theta_{1:T}|\mathbf{X}^{1:T},\mathbf{w}} \nonumber\\
& -\frac{T-1}{2}\log{|2\pi \mathbf{Q}^{\ast}|} 
- \frac{1}{2}\sum\limits _{t=2}^{T}\left\langle\left(\boldsymbol{\theta}_{t}-\boldsymbol{\theta}_{t-1}\right)^{\prime}\mathbf{Q}^{\ast-1}\left(\boldsymbol{\theta}_{t}-\boldsymbol{\theta}_{t-1}\right)\right\rangle_{\boldsymbol\theta_{1:T}|\mathbf{X}^{1:T},\mathbf{w}}.
\end{align}
By considering derivatives of this equation w.r.t. the hypermarameters, we obtain their update rules. The precision $\lambda^\star\mathbf{I} (= \mathbf{Q^{\ast-1}})$ is updated as
\begin{equation}
\lambda^{\ast}=\frac{1}{(T-1)d}{\rm tr}\left[ \sum_{t=2}^T  \left\langle(\boldsymbol{\theta}_t-\boldsymbol{\theta}_{t-1})(\boldsymbol{\theta}_t-\boldsymbol{\theta}_{t-1})^\prime\right\rangle_{\boldsymbol\theta_{1:T}|\mathbf{X}^{1:T},\mathbf{w}}\right],
\end{equation}
where $d$ is the dimension of vector $\boldsymbol\theta_t$. The initial mean is optimized by $\boldsymbol\mu^\ast=\langle\boldsymbol{\theta}_{1} \rangle_{\boldsymbol\theta_{1:T}|\mathbf{X}^{1:T},\mathbf{w}}$. Here the key step is to develop an algorithm that constructs the posterior density of Eq~\ref{eq:smoother}. This is done by the forward and backward recursive Bayesian algorithms. Below we review this method followed by introduction of the approximations that make the method applicable to larger number of neurons. 

\bigskip \noindent 
\textbf{Recursive estimation of dynamic neural interactions} 
The estimation of the latent process is achieved by forward filtering and then backward smoothing algorithms. In the filtering algorithm, we sequentially estimate the state of population activity at time bin $t$ given the data up to time $t$. This estimate is given by the recursive Bayesian formula
\begin{align}\label{eq:posterior}
p(\boldsymbol{\theta}_t|\mathbf{X}^{1:t},\mathbf{w})
= \frac{p(\mathbf{X}^{t} | \boldsymbol{\theta}_t) p(\boldsymbol{\theta}_t | \mathbf{X}^{1:t-1},\mathbf{w}) } {p(\mathbf{X}^{t} | \mathbf{X}^{1:t-1}, \mathbf{w})}.
\end{align}
where $p(\mathbf{X}^{t} | \boldsymbol{\theta}_t)$ is obtained from the observation model. The second term in the numerator $p(\boldsymbol{\theta}_t|\boldsymbol{X}^{1:t-1},\mathbf{w})$ is called the one-step prediction density. It is computed using the state model and the filter density at the previous time bin via the Chapman-Kolmogorov equation, 
\begin{equation}\label{eq:chapmankolmogorov}
p(\boldsymbol{\theta}_t|\boldsymbol{X}^{1:t-1},\mathbf{w}) = \int p(\boldsymbol{\theta}_t| \boldsymbol{\theta}_{t-1},\mathbf{w}) p(\boldsymbol{\theta}_{t-1}|\boldsymbol{X}^{1:t-1},\mathbf{w}) d\boldsymbol{\theta}_{t-1}.
\end{equation}
Thus the filter density (Eq~\ref{eq:posterior}) can be recursively computed for $t=2,\ldots,T$ using Eq~\ref{eq:chapmankolmogorov}, given observation and state models as well as an initial distribution of the one-step prediction density at time $t=1$. Note that the initial one-step prediction density was specified as $p(\boldsymbol\theta_{1}) = \mathcal{N}(\boldsymbol\mu,\boldsymbol\Sigma)$. This distribution dictates the density of the state at the initial time step without observing neural activity. 

The approximate nonlinear recursive formulae were developed by approximating the posterior density (Eq~\ref{eq:posterior}) with a Gaussian distribution \cite{fahrmeir1992posterior,brown1998statistical}. Let us assume that the filter density at time $t-1$ is given by a Gaussian distribution with mean $\boldsymbol{\theta}_{t-1|t-1}$ and the covariance matrix $\mathbf{W}_{t-1|t-1}$. The subscript $t-1 \vert t-1$ means the estimate at time $t-1$ (left) given the data up to time bin $t-1$ (right). Because the state model (Eq~\ref{eq:statemodel}) is also Gaussian, the Chapman-Kolmogorov equation yields the one-step prediction density that is a Gaussian distribution with mean $\boldsymbol\theta_{t|t-1} = \boldsymbol\theta_{t-1|t-1}$ and covariance $\mathbf{W}_{t|t-1} = \mathbf{W}_{t-1|t-1} + \mathbf{Q}$. We then obtain the following log posterior density (Eq~\ref{eq:posterior}), 
\begin{align}\label{eq:log_posterior}
\log p(\boldsymbol{\theta}_t|\mathbf{X}^{1:t},\mathbf{w}) = 
& \sum_{r=1}^{R} \left[ \boldsymbol{\theta}_t^\prime \mathbf{F}(\mathbf{X}^{t,r}) - \psi(\boldsymbol{\theta}_t) \right] \nonumber \\
& - \frac{1}{2} (\boldsymbol{\theta}_t - \boldsymbol{\theta}_{t|t-1})' \mathbf{W}^{-1}_{t|t-1} (\boldsymbol{\theta}_t - \boldsymbol{\theta}_{t|t-1}) + \mathrm{const.}
\end{align}
Here we approximate the posterior density by a Gaussian distribution (the Laplace approximation). We identify the mean of this distribution with the MAP estimate:
\begin{equation} \label{eq:map_posterior}
\boldsymbol\theta_{t|t} = \mbox{argmax}_{\boldsymbol\theta_t} \log p(\boldsymbol{\theta}_t|\mathbf{X}^{1:t},\mathbf{w}).
\end{equation}
This solution is called a filter mean. It may be obtained by gradient ascent algorithms such as the conjugate gradient algorithm and the Broyden-Fletcher-Goldfarb-Shanno (BFGS) algorithm. These algorithms use the gradient
\begin{equation}\label{eq:dlog_posterior}
\frac{\partial \log p( \boldsymbol{\theta}_t|\mathbf{X}^{1:t},\mathbf{w})}{\partial \boldsymbol{\theta}_t}  = \sum_{r=1}^{R} \left[ \mathbf{F}(\mathbf{X}^{t,r}) - \boldsymbol{\eta}_t \right] - \mathbf{W}^{-1}_{t|t-1} (\boldsymbol{\theta}_t - \boldsymbol{\theta}_{t|t-1}). 
\end{equation}
Here we define the expectation parameters $\boldsymbol{\eta}_t$ as 
\begin{equation}\label{eq:expectation}
\boldsymbol{\eta}_t \equiv \frac{\partial \psi(\boldsymbol{\theta}_t)}{\partial \boldsymbol{\theta}_t } = \langle \mathbf{F}(\mathbf{x}) \rangle_{\boldsymbol\theta_t},
\end{equation}
where $\langle \mathbf{x} \rangle_{\boldsymbol\theta_t}$ is the expectation of $\mathbf{x}$ with respect to $p(\mathbf{x}\vert\boldsymbol\theta_t)$. This expectation needs to be computed repeatedly in the gradient algorithms. The covariance matrix of the approximated Gaussian distribution is computed from the Hessian of the log posterior evaluated at the MAP estimate:
\begin{align}\label{eq:dd_posterior}
\mathbf{W}_{t|t}^{-1} = 
& - \left. \frac{\partial^2 \log p(\boldsymbol{\theta}_t|\mathbf{X}^{1:t},\mathbf{w}) }{\partial \boldsymbol{\theta}_t \partial \boldsymbol{\theta}_t^\prime} \right|_{\boldsymbol{\theta}_{t|t} } \nonumber\\
 = & R\mathbf{G}_t + \mathbf{W}_{t|t-1}^{-1}.
\end{align}
$\mathbf{G}_t$ is the Fisher-information matrix:
\begin{align}\label{eq:fisher_matrix}
\mathbf{G}_t \equiv 
\left. \frac{\partial \psi(\boldsymbol{\theta}_t)}{\partial \boldsymbol{\theta}_t \partial \boldsymbol{\theta}_t ^\prime } \right |_{\boldsymbol{\theta}_t = \boldsymbol{\theta}_{t|t}}
= \langle \mathbf{F}(\mathbf{x})\mathbf{F}(\mathbf{x})^\prime\rangle_{\boldsymbol{\theta}_{t|t}} - \langle\mathbf{F}(\mathbf{x})\rangle_{\boldsymbol{\theta}_{t|t}}\langle\mathbf{F}(\mathbf{x})\rangle_{\boldsymbol{\theta}_{t|t}}^\prime .
\end{align}
The expectations are taken by $p(\mathbf{x} |\boldsymbol{\theta}_{t|t})$. Note that we initially assumed that the filter density at previous time step is a Gaussian distribution when computing the Chapman-Kolmogorov equation. By the Laplace approximation, this assumption is fulfilled in the next time step. Additionally we assumed that the initial distribution of the state variables is Gaussian. Thus we obtain an approximate nonlinear recursive filter that is consistent across the iterations. 

Once the approximate filter density is constructed for $t=1,\ldots,T$ , the backward smoothing algorithm is applied to obtain the smoothed posterior density of the state variable at time $t$\cite{kitagawa1987non,brown1998statistical},
\begin{equation}
p(\boldsymbol{\theta}_{t}|\boldsymbol{X}^{1:T},\mathbf{w})=p(\boldsymbol{\theta}_{t}|\boldsymbol{X}^{1:t},\mathbf{w})\int\frac{p(\boldsymbol{\theta}_{t+1}|\boldsymbol{X}^{1:T},\mathbf{w}) p(\boldsymbol{\theta}_{t+1}|\boldsymbol{\theta}_{t},\mathbf{w})}{p(\boldsymbol{\theta}_{t+1}|\boldsymbol{X}^{1:t},\mathbf{w})}d\boldsymbol{\theta}_{t+1}.
\end{equation}for $t=T,\ldots,1$. In practice, the following fixed interval smoothing algorithm \cite{brown1998statistical} provides the smoothed MAP estimate  $\boldsymbol\theta_{t\vert T}$ and smoothed covariance $\mathbf{W}_{t|T}$ of the posterior distribution
\begin{align}
\boldsymbol{\theta}_{t|T} 
&= \boldsymbol{\theta}_{t|t} + \mathbf{A}_t(\boldsymbol{\theta}_{t+1|T} - \boldsymbol{\theta}_{t+1|t}),
\\
\mathbf{W}_{t|T} &= \mathbf{W}_{t|t} + \mathbf{A}_t(\mathbf{W}_{t+1|T} - \mathbf{W}_{t+1|t})\mathbf{A}_t^\prime,
\end{align}
where $\mathbf{A}_t= \mathbf{W}_{t|t}\mathbf{W}_{t+1|t}^{-1}$. In addition, the posterior covariance matrix between state variables at time $t$ and $t-1$ is obtained as $\mathbf{W}_{t-1,t|T}=\mathbf{A}_{t-1} \mathbf{W}_{t|T}$ \cite{de1988covariances}. This procedure constructs the smoother posterior density of the latent process (Eq~\ref{eq:smoother}) by approximating it as a Gaussian process of length $N(N+1)/2 \times T$ with mean $(\boldsymbol{\theta}_{1|T}^{\prime},\boldsymbol{\theta}_{2|T}^{\prime},\ldots,\boldsymbol{\theta}_{T|T}^{\prime})$ and a block tridiagonal covariance matrix whose block diagonal is given by $\mathbf{W}_{t|T}$ (for $t = 1,\dots,T$), and block off-diagonals are given by $\mathbf{W}_{t-1,t|T}$ (for $t = 2,\dots,T$).

\subsection*{Approximation methods for large-scale analysis}
\par
\subsubsection*{Approximate estimate of filter mean by pseudolikelihood method} To obtain the filter estimate using iterative gradient ascent methods, the gradient (Eq~\ref{eq:dlog_posterior}) needs to be evaluated at each iteration. This requires computation of the expectations (Eq \ref{eq:expectation}) by summing over all $2^N$ states the network can realize. This is infeasible for a large network size $N$. Thus the method introduced in the previous subsection was limited to $N\leq 15$. However, the \emph{pseudolikelihood} method \cite{besag1975statistical, hofling2009estimation,kolar2010estimating} has been shown to estimate with reasonable accuracy the interactions without requiring evaluation of the expectations. Here we incorporate it into the sequential Bayesian estimation framework. 
\par
The pseudolikelihood approximates the likelihood of the joint activity of neurons by a product of conditional likelihoods of each neuron given the activity of the others. Let the activity of neurons except neuron $n$ be $\mathbf{x}_{\backslash n} =(x_1,...,x_{n-1},x_{n+1},...,x_{N})^\prime$; and $f^n_t(\mathbf{x}_{\backslash n})=\boldsymbol{\theta}_t^\prime\mathbf{F}(x_n=1,\mathbf{x}_{\backslash n})$. Then the pseudolikelihood is given by
\begin{eqnarray}\label{eq:pseudolikelihood}
\prod_{r=1}^R\tilde{p}(\mathbf{X}^{t,r}|\boldsymbol{\theta}_t) = \prod_{r=1}^R\prod_{n=1}^N p\left(X^{t,r}_n|\mathbf{X}^{t,r}_{\backslash n},\boldsymbol{\theta}_t\right) = \prod_{r=1}^R\prod_{n=1}^N\frac{\exp\left(X^{t,r}_nf^n_t\left(\mathbf{X}^{t,r}_{\backslash n}\right)\right)}{1+\exp\left(f^n_t\left(\mathbf{X}^{t,r}_{\backslash n}\right)\right)}.
\end{eqnarray}
Note that the log partition function does not appear in Eq \ref{eq:pseudolikelihood}. Replacing the likelihood in Eq~\ref{eq:posterior} with Eq~\ref{eq:pseudolikelihood} yields
\begin{align}\label{eq:pseudomap}
\log p(\boldsymbol{\theta}_t\vert \mathbf{X}^{1:T},\mathbf{w})  \approx & \sum_{r=1}^R\sum_{n=1}^N\left[X^{t,r}_nf_t^n\left(\mathbf{X}^{t,r}_{\backslash n}\right) - \log\left(1+\exp\left(f_t^n\left(\mathbf{X}^{t,r}_{\backslash n}\right)\right)\right)\right] \nonumber\\
& -\frac{1}{2}(\boldsymbol\theta_{t} - \boldsymbol\theta_{t\vert t-1})^\prime \mathbf{W}_{t\vert t-1}^{-1}(\boldsymbol\theta_{t} - \boldsymbol\theta_{t\vert t-1}) + \mbox{const.}
\end{align}
The derivative of this approximated filter density results in
\begin{align}\label{eq:pseudoderivative}
\frac{\partial \log p(\boldsymbol{\theta}_t\vert \mathbf{X}^{1:T},\mathbf{w})}{\partial \boldsymbol{\theta}_t}  \approx & \sum_{r=1}^R\sum_{n=1}^N\left[ \left(X^{t,r}_n - \tilde{\eta}^{t,r}_n\right)\frac{\partial f^n_t(\mathbf{X}^{t,r})}{\partial \boldsymbol{\theta}_t} \right] -  \mathbf{W}^{-1}_{t|t-1} (\boldsymbol{\theta}_t - \boldsymbol{\theta}_{t|t-1}),
\end{align}
where $\tilde{\eta}^{t,r}_n=\left\langle x^t_n \vert \mathbf{X}_{\backslash n}^{t,r} \right\rangle_{\boldsymbol{\theta}_t}$, i.e., the expectation of $x_n^t$ being $1$ given the activity of the other neurons. Using this gradient in the same gradient ascent algorithms as before we obtain the approximate mean $\boldsymbol\theta_{t|t}$ of the filter density. 
\par
\subsubsection*{Approximation of the filter covariance} The pseudolikelihood can provide the approximate mode of the filter density (Eq \ref{eq:map_posterior}). However, to perform the sequential estimation, we need in addition the filter covariance matrix (Eq~\ref{eq:dd_posterior}). This requires to compute the Fisher information matrix (Eq~\ref{eq:fisher_matrix}, i.e., the Hessian of the observation model at the filter mean $\boldsymbol\theta_{t|t}$). To compute the Fisher information matrix, not only the first and second order but also the third and fourth order expectation parameters need to be evaluated at the filter mean parameters. In order to avoid computing the higher-order expectation parameters and to reduce the computational cost of the matrix inversion, we approximate it by a diagonal matrix. The diagonal is composed of the first and second order expectation parameters $\lbrace \eta_i^{t|t} \rbrace, \lbrace \eta_{ij}^{t|t} \rbrace$, where the expectations parameters are defined as $\eta_i^{t|t} \equiv \langle x_i\rangle_{\boldsymbol\theta_{t|t}}$ and $\eta_{ij}^{t|t} \equiv \langle x_i x_j\rangle_{\boldsymbol\theta_{t|t}}$. Here we test two different approximation methods to obtain these marginals. One is the \emph{Bethe approximation} \cite{yedidia2003understanding} and the other the mean-field \emph{Thouless-Anderson-Palmer} (\emph{TAP}) approach \cite{thouless1977solution}.

\bigskip \noindent 
\textbf{Bethe approximation} The Bethe approach approximates a probability distribution by assuming that it factorizes into its pairwise marginals. Hence, the approximated joint distribution writes as
\begin{equation}\label{eq:bethe_approx}
p(\mathbf{x}|{\boldsymbol\theta}_{t|t}) \approx \frac{\prod_{i,j>i} q_t(x_i,x_j)}{\prod_i q_t(x_i)^{(N-1)-1}} := q_t (\mathbf{x}),
\end{equation}
where $q$ are so-called \emph{beliefs} \cite{yedidia2001idiosyncratic} that approximate the marginals of the underlying distribution $p$. Note that for any acyclic graph this yields the true joint distribution. However, here the observation model (Eq~\ref{eq:observationmodel}) is a fully connected graph and hence the Bethe approximation ignores all cycles. Realizing that the beliefs have to fulfill constraints ($\sum_{x_j}q_t(x_i,x_j)=q_t(x_i)$ and $\sum_{x_i}q_t(x_i)=1$) one can write the problem as a Lagrangian that has to be minimized. This allows to derive a dual representation of the marginals (in terms of the Lagrangian multipliers), which in turn allows to derive messages that are sent from one belief to another. Propagating this beliefs through the Markov field yields the belief propagation algorithm (BP) \cite{yedidia2003understanding}. While BP is relatively fast in obtaining the expectation values, it is not guaranteed to converge to an unique solution. This guarantee is provided by the alternative concave-convex procedure (CCCP) \cite{yuille2002cccp}. CCCP also starts from the same Lagrangian, but updates the beliefs and Lagrangian multipliers in an alternating manner. This more strict procedure comes with the disadvantage that it is much slower than BP. Therefore, here the two algorithms are combined to a \emph{hybrid method}, where BP is utilized primarily and the algorithm falls back to CCCP, when BP does not converge. For more details on the Bethe approximation, see \nameref{SI:text 1}.
\par
The estimation of the log partition function for the Bethe approximation is simply computed by the negative logarithm of the approximated probability (Eq~\ref{eq:bethe_approx}) that all neurons are silent, i.e.,
\begin{equation}\label{eq:bethe_psi_main}
\psi_t \approx -\log q_t(\boldsymbol{0}).
\end{equation}

\bigskip \noindent 
\textbf{TAP approximation} The TAP approximation of the expectation parameters $\boldsymbol\eta_{t\vert t}$ given the natural parameters $\boldsymbol\theta_{t\vert t}$ (\emph{forward-problem}) can be derived in multiple ways \cite{opper2001advanced,roudi2009statistical}, but here we follow \cite{tanaka1998mean,tanaka1999theory} that use the so-called ``Plefka expansion''. The following formulae and their derivation are revised for binary variables $x_i \in\lbrace 0,1 \rbrace$ instead of $\lbrace -1,1 \rbrace$. See \nameref{SI:text 2} for more details. The method constructs a new free energy as a function of the mixture coordinates $(\{ \eta_i^{t|t} \}, \{\theta_{ij}^{t|t}\})$ by the Legendre transformation of the log partition function $\psi_t$ as $\sum_{i=1}^N{\theta_i^{t|t} \eta_i^{t|t}} - \psi_t$. Then this function is approximated by a second-order expansion around the independent model assuming weak pairwise interactions. This results in the approximate log partition function,
\begin{align} \label{eq:psi_TAPapprox_main}
\psi_t \approx & \sum_{i=1}^N\theta_i^{t|t} \eta_i^{t|t} - \sum_{i=1}^N \left( \eta_i^{t|t} \log \eta_i^{t|t} + (1-\eta_i^{t|t})\log(1-\eta_i^{t|t}) \right) +\frac{1}{2}\sum_{j\neq i} \theta_{ij}^{t|t} \eta_i^{t|t} \eta_j^{t|t} \nonumber\\
& +\frac{1}{8}\sum_{j\neq i} \left(\theta_{ij}^{t|t}\right)^2 \left(\eta_i^{t|t} - (\eta_i^{t|t})^2\right) \left(\eta_j^{t|t} - (\eta_j^{t|t})^2\right).
\end{align}
Here we extended the definition of interaction parameters as $\theta_{ii}^{t|t}=0$ and $\theta_{ij}^{t|t}=\theta_{ji}^{t|t}$. At the independent model, the values for the expectations can be computed and the expansion yields correction terms for the non-zero $\theta_{ij}^{t|t}$. Since derivatives of the new free energy based on the mixture coordinates w.r.t. $\{\eta_i^{t|t}\}$ yield the first order parameters $\{\theta_i^{t|t}\}$, we obtain the following self-consistent equations:
\begin{equation}
\theta_i^{t|t} = \log \left(\frac{\eta_i^{t|t}}{1-\eta_i^{t|t}}\right) - \sum_{j\neq i} \theta_{ij}^{t|t}\eta_j - \frac{1}{2}\sum_{j\neq i} \left(\theta_{ij}^{t|t}\right)^2 \left(\frac{1}{2}-\eta_i^{t|t}\right) \left(\eta_j^{t|t}-(\eta_j^{t|t})^2\right),
\label{eq:tap_2} 
\end{equation}
for $i,j=1,\ldots,N$. Solving this equations yields the first order expectations which can be used to estimate the log partition function (Eq~\ref{eq:psi_TAPapprox_main}). 
\par
Furthermore, from the relation $\frac{\partial \theta_i^{t|t}}{\partial \eta_j^{t|t}} = [\mathbf{G}_t^{-1}]_{ij}$ we obtain 
\begin{equation}
[\mathbf{G}_t^{-1}]_{ij} = \frac{1}{\eta_i^{t|t} (1 - \eta_i^{t|t})} \delta_{ij} - \theta_{ij}^{t|t} - \left(\theta_{ij}^{t|t}\right)^2 \left(\frac{1}{2} - \eta_i^{t|t}\right) \left(\frac{1}{2} - \eta_j^{t|t}\right). \label{eq:tap_1}
\end{equation}
Here $\delta_{ij}$ is the Kronecker delta function, which is $1$ for $i=j$ and $0$ otherwise. To obtain the second order expectation parameters, we calculate and then invert the $N \times N$ matrix obtained by Eq~\ref{eq:tap_1}, and approximate it as the Fisher information matrix for $\{\theta_i\}$ given in Eq~\ref{eq:fisher_matrix} to obtain the second order expectation parameters by $\eta_{ij}^{t|t}=[\mathbf{G}_t]_{ij}+\eta_i^{t|t} \eta_j^{t|t}$ \cite{tanaka1998mean}. 

\bigskip \noindent 
\textbf{Approximate marginal likelihood} Because the TAP and Bethe approximations provide estimates of the log partition function $\psi_t$, we are able to evaluate the approximation of the marginal likelihood (Eq~\ref{eq:Q-function}), and the EM-algorithm for the state-space model can be run until it converges. The approximate marginal likelihood is obtained as (see also \cite{shimazaki2012state})
\begin{align}
l(\mathbf{X}^{1:T} |\mathbf{w})
 = &\sum_{t=1}^{T}\log\int p(\mathbf{X}^{t}|\boldsymbol{\theta}_{t})p(\boldsymbol{\theta}_{t}|\mathbf{X}^{1:t-1},\mathbf{w})d\boldsymbol{\theta}_{t} \nonumber \\
 \approx & 
 \sum_{t=1}^{T} \sum_{r=1}^{R} 
\left[ \boldsymbol{\theta}_{t|t}^\prime \mathbf{F}(\mathbf{X}^{t,r})
-\psi_t\left(\boldsymbol{\theta}_{t|t}\right)\right] \nonumber \\
 & -\frac{1}{2}\sum_{t=1}^{T} \left(\boldsymbol{\theta}_{t|t}-\boldsymbol{\theta}_{t|t-1}\right)^{\prime} \mathbf{W}_{t|t-1}^{-1} \left(\boldsymbol{\theta}_{t|t}-\boldsymbol{\theta}_{t|t-1}\right) \nonumber \\
 & +\frac{1}{2}\sum_{t=1}^{T}\left(\log{\det W_{t|t}}-\log{\det W_{t|t-1}}\right),
 \label{eq:approx_marginal_loglikelihood}
\end{align}
where $p(\boldsymbol{\theta}_{t}|\mathbf{X}^{1:0},\mathbf{w})$ indicates a prior of the initial distribution $\mathcal{N}(\boldsymbol\mu,\boldsymbol\Sigma)$. Similarly, we use $\boldsymbol\theta_{1|0} = \boldsymbol\mu$ and $\mathbf{W}_{1|0} = \boldsymbol\Sigma$. Here the integral with respect to $\boldsymbol \theta_t$ at the first equality is approximated as an integral of a Gaussian function, using up to the quadratic information around its mode (the Laplace approximation). From Eqs~\ref{eq:log_posterior} and \ref{eq:map_posterior}, it turns out that the mean and covariance of the filter density provide this information.

\section*{Results}
\begin{figure}[t]
\begin{center}
\includegraphics[width=\textwidth]{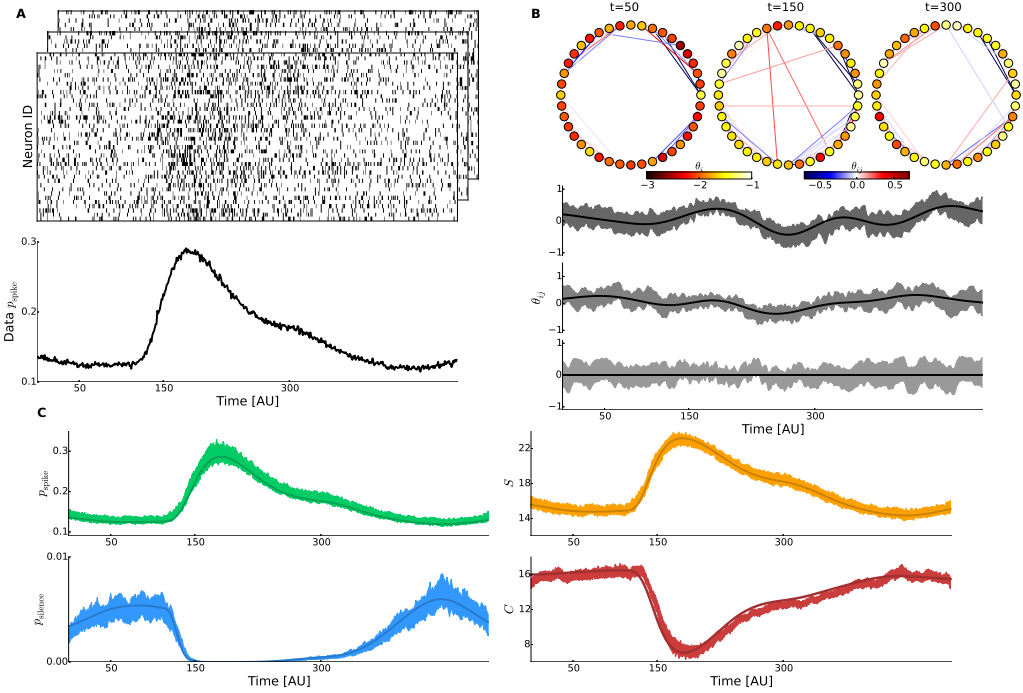}
\end{center}
\caption{ {\bf Approximate inference of dynamic neural interactions and macroscopic network properties.} Analysis on simulated spike data of $40$ neurons.
\textbf{A} Top: Simultaneous spiking activity of 40 neurons that are repeatedly simulated 500 times (here only 3 trials are visualized). The data is sampled from a time-dependent model of a neural population (Eq~\ref{eq:observationmodel}). The time-varying parameters are chosen such that neurons' spike probability resembles evoked activity in response to stimulus presentation to an animal. The neural interactions are assumed to smoothly change irrespective of the firing rates. See the main text for details. Bottom: Empirical spike probability over time, averaged over trials and neurons. 
\textbf{B} Top: Estimated network states at $t=50,150,300$ by the pseudolikelihood-Bethe approximation method. Neurons are represented by nodes whose colors respectively indicate a value of the smoothed estimate of $\theta_i^{t}$ (for $i=1,\ldots,40$). Links are color-coded according to estimated strength of the interaction $\theta_{ij}^{t}$ between connected nodes (positive or negative interactions are marked in red or blue, respectively). Only significant edges are displayed, where the corresponding $\theta_{ij}^t$ has a $98\%$ credible interval that does not include $0$. Bottom: Dynamics of 3 exemplary interaction parameters, $\theta_{ij}^t$. The lines denote the ground truth from which the binary data are sampled. The shaded areas are $98\%$ credible intervals. 
\textbf{C} Estimated population rate (top left). Probability that all neurons are silent (bottom left). Entropy (top right) and heat capacity (bottom right) of the neural population. In all panels, shaded areas indicate $1\%$ and $99\%$ quantiles obtained by resampling the natural parameters from the fitted smoothed distribution. Solid lines represent ground truth computed from the underlying network model.}\label{fig:bethe_example}
\end{figure}

\textbf{Model fit to simulated data} In the following subsections, we demonstrate the fit of the state-space model of neural population activity to artificially generated data of $40$ neurons with dynamic couplings for $T=500$ time bins. To be able to compare it to the ground truth we construct $4$ populations each consisting of $10$ neurons. Individual parameters $\boldsymbol\theta_{1:T}$ of the underlying submodels are generated as smooth independent Gaussian processes, where the mean for the first order parameters $\theta_i^t$ increases at $t=100$ and then decreases more slowly shortly after that. The interaction parameters $\theta_{ij}^t$ are generated as Gaussian processes whose mean is fixed at $0$. In total, $500$ trials of spike data are sampled from this generative model. Note that the sampled individual parameters differ and vary over time although we use homogeneous means. The increase of the mean for $\theta_i^t$ increases spiking probability followed by a decrease back to baseline (Fig~\ref{fig:bethe_example}\textbf{A}). In the resulting data neurons spike with time averaged probabilities ranging from $0.10$ up to $0.21$. Supposing bin width $\Delta = 10\ \mathrm{ms}$ these are in a physiologically reasonable range. This exemplary scenario may mimic a population that independently receives an external input elicited by e.g., a sensory stimulus. For details of the generation of the data see \nameref{SI:text 3}.

\par 
Next we fit the state-space model of neural population activity to the generated data with the combination of pseudolikelihood and Bethe approximation. This combination is chosen for the demonstration because it provides the best estimates of the underlying model as we will assess  later in this section. Top panel of Fig~\ref{fig:bethe_example}\textbf{B} shows snapshots of the smoothed estimates of the inferred network at different time points ($t=50, 150, 300$). The color of the nodes indicate the smoothed estimates of the first order parameters $\theta_{i}^{t|T}$ and the one of the edges interactions $\theta_{ij}^{t|T}$. Visual inspection of the fitted network suffices to identify that there are $4$ independent subpopulations of correlated neurons (one in each quadrant). To check whether the inferred changes over time match those of the underlying generative model, credible intervals of three fitted couplings are compared with their underlying values (Fig~\ref{fig:bethe_example}\textbf{B} Bottom). The fit follows the dynamics, and correctly identifies the parameter that is constantly $0$ (the lowest panel).

\bigskip \noindent 
\textbf{Estimating macroscopic properties of the network} One of the main motives to model joint activities of a large population of neurons is to assess macroscopic properties of the network in a time-dependent manner with credible intervals. The macroscopic measures obtained for this example are shown in Fig~\ref{fig:bethe_example}\textbf{C}, and in the following we introduce them one by one.
\par
The first and simplest macroscopic property shown in the top left panel of Fig~\ref{fig:bethe_example}\textbf{C} is the probability of spiking in a network (population spike rate). We define it as
\begin{equation}\label{eq:spikeprob}
p_{\rm{spike}}(t) = \frac{1}{N}\sum_{i=1}^N \eta_{i}^{t},
\end{equation}
where $\eta_{i}^{t}$ is the spike rate of $i$th neuron at time $t$. Considering the smoothed estimate $\eta_{i}^{t} = \eta_{i}^{t|T}$, the method recovers correctly the empirical rate obtained from the data (Fig~\ref{fig:bethe_example}\textbf{A} Bottom). The shaded area in the panel indicates the 98\% credible interval of the population spike rate obtained by resampling the natural parameters from the smoothed posterior density $100$ times at each bin. The underlying spike probability for $N=40$ neurons is obtained by calculating the marginals $\eta_i^t$ independently for each subpopulation and averaging over all neurons.
\par 
Next from the state-space model of neural population activity one can estimate the probability of simultaneous silence (i.e., the probability that no neuron elicits a spike, Fig~\ref{fig:bethe_example}\textbf{C} bottom left)
\begin{equation}\label{eq:probsilence}
p_{\rm{silence}}(t) = \exp(-\psi_t).
\end{equation}
The approximation methods allow us to evaluate the log partition function $\psi_t$ (Eqs \ref{eq:bethe_psi_main} and \ref{eq:psi_TAPapprox_main}). Here we use smoothed estimates to compute the log partition function. Thus we immediately obtain the probability of simultaneous silence. The expected simultaneous silence for $N=40$ neurons is obtained as multiplication of the silence probabilities of the $4$ subpopulations. 

\par 
The entropy of the network (i.e., expectation of the information content, $\langle - \log p(\mathbf{x}|\boldsymbol\theta_t) \rangle_{\boldsymbol\theta_t}$) can be also calculated from the model as
\begin{equation}\label{eq:entropy}
S(t) = -\boldsymbol\theta_t^\prime\boldsymbol\eta_t + \psi_t. 
\end{equation}
Estimation of this information theoretic measure allows us to quantify the amount of interactions in the network by comparing the pairwise model to the independent one (see following analyses and Eq~\ref{eq:explained_entropy}). Since it is an extensive quantity, the entropy of $N=40$ neurons is obtained by addition of the entropies from the $4$ independent subpopulations. The entropy increases while the individual activity rates of neurons also increases (Fig~\ref{fig:bethe_example}\textbf{C} top right). 
\par 
The last measure shown in the bottom right panel of Fig~\ref{fig:bethe_example}\textbf{C} is the heat capacity, or sensitivity, of the system. It is the variance of information content: $C(t)=\langle \{-\log p(\mathbf{x}|\boldsymbol\theta_t)\}^2 \rangle_{\boldsymbol\theta_t} - \{ \langle -\log p(\mathbf{x}|\boldsymbol\theta_t) \rangle_{\boldsymbol\theta_t} \}^2$, where the brackets indicate expectation by $p(\mathbf{x}|\boldsymbol\theta_t)$. It is also the variance of the Hamiltonian $- \boldsymbol{\theta}_t^\prime\mathbf{F}(\mathbf{x})$. Thus we can obtain it by introducing a nominal dual parameter $\beta$ to the Hamiltonian in the model, assuming that it is $1$ for real data. The log partition function of the augmented model is 
\begin{equation}
\psi_t(\beta) = \log\sum_{\mathbf{x}}\exp(\beta \, \boldsymbol{\theta}_t^\prime\mathbf{F}(\mathbf{x})).
\end{equation}
The variance of Hamiltonian is given as the Fisher information w.r.t. $\beta$, i.e., the second derivative of the log partition function. This allows us to use the approximate $\psi_t$ to assess the heat capacity. Then we further approximate the second derivative by its discrete version
\begin{equation}\label{eq:sensitivity}
C(t) = \left. \frac{\partial^2 \psi_t}{\partial \beta^2} \right |_{\beta=1} \approx \frac{\psi_t(1+\epsilon) - 2\psi_t(1) + \psi_t(1-\epsilon)}{\epsilon^2},
\end{equation}
and $\epsilon$ is chosen to be $10^{-3}$. The heat capacity measures sensitivity of the network, namely how much the network activity changes due to subtle changes in its network configuration (i.e., to changes of the $\boldsymbol{\theta}_t$ parameters). Networks with higher sensitivity are more responsive to changes than those with lower sensitivity. Similarly to the entropy, the heat capacity is an extensive quantity. For the simulated data, the heat capacity decreases while activity rates of neurons are increased (Fig~\ref{fig:bethe_example}\textbf{C} bottom right).    

\begin{figure}[tb]
\centering
\includegraphics[width=1.\textwidth]{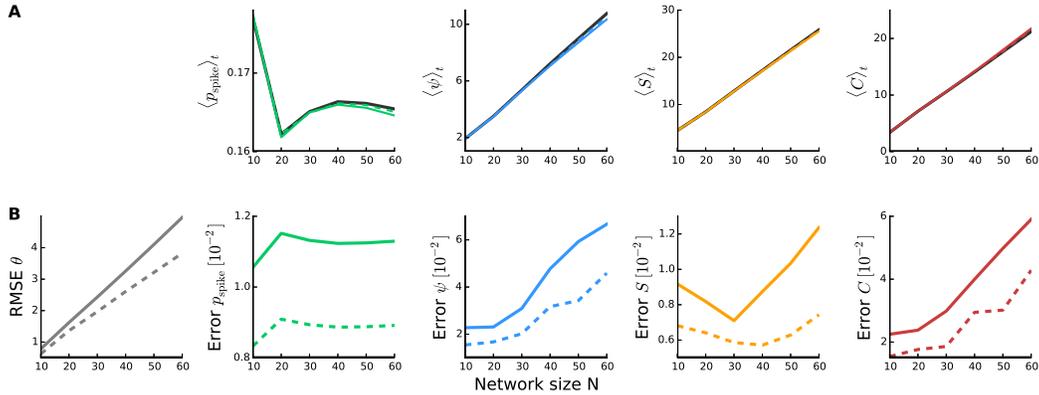}
\caption{\textbf{Approximation error and network size.} Error analysis on networks consisting of subpopulations with $10$ neurons, constructed by the same procedure as in Fig~\ref{fig:bethe_example}. \textbf{A}: The average value of the macroscopic properties over time as a function of network size. Black line is the true value, while colored lines show the estimated ones (solid line fit with $500$ trials and dashed with $1000$ trials) \textbf{B}: The corresponding errors (only for $\boldsymbol\theta_t$ the RMSE is shown) for $500$ trials (solid) and $1000$ trials (dashed).}\label{fig:bethe_large}
\end{figure}

\bigskip \noindent 
\textbf{Assessment of fitting error with different network sizes and amount of data} Next we examine the goodness-of-fit of the model fitted by the pseudolikelihood and Bethe approximation methods. In particular, we ask how the fitting performance changes with increasing network size. For this reason we generated $6$ dynamic models for populations of $10$ neurons as described previously ($500$ time bins, $500$ trials). Then we construct smaller or larger populations by concatenating the independent groups. The model is fitted by the pseudolikelihood and Bethe approximation methods to the first subnetwork, then two subnetworks, and so on, until we fit the model to a network containing $60$ neurons composed of $6$ independent groups. We obtain estimates of the macroscopic measures from the smoothed estimates of the model parameters at each time bin. Figure \ref{fig:bethe_large}\textbf{A} shows values of these measures averaged over time. The results show extensive properties of macroscopic measures (except for the population spike rate), and that the estimates may slightly deviate for larger number of neurons. 

\par 
To assess quality of the fit, first the root mean squared error (RMSE) for the natural parameters averaged across time bins is calculated
\begin{equation}\label{eq:rmise}
\mbox{RMSE}(\boldsymbol\theta_{t|T}) = \sqrt{\frac{1}{T}\sum_{t=1}^T \left\Vert\boldsymbol\theta_{t|T} - \boldsymbol\theta_{t}\right\Vert^2 },
\end{equation}
where $\boldsymbol\theta_{t|T}$ is the smoothed estimate of the underlying model $\boldsymbol\theta_t$. $\Vert \boldsymbol{v} \Vert$ denotes the $L2$-norm of vector $\boldsymbol{v}$. For the data sets with $500$ trials, the RMSE increases linearly with network size (Fig~\ref{fig:bethe_large}\textbf{B} Left). Furthermore, the error for the macroscopic measures is assessed by
\begin{equation}\label{eq:error}
\mbox{Error}[f(\boldsymbol\theta_{t|T})] = \frac{\mbox{RMSE}(f(\boldsymbol\theta_{t|T}))}{\frac{1}{T}\sum_{t=1}^T f(\boldsymbol\theta_{t})},
\end{equation}where $f(\boldsymbol\theta_{t|T})$ is any function of the macroscopic measures. The RMSE is defined similarly to Eq~\ref{eq:rmise} while substituting the parameters $\boldsymbol\theta_{t|T}$ by the function $f(\boldsymbol\theta_{t|T})$. Besides the population rate these errors also increase as the network size increases (Fig~\ref{fig:bethe_large}\textbf{B}). We observe non-monotonic behavior in some of the macroscopic properties (e.g., average spike rate and the entropy's error), which can be explained by fluctuations from the data generation process.
\par 
To understand whether these errors increase primarily due to the approximation methods used for the fit or because of the finite amount of data, the fit is repeated but now to spiking data with $1000$ trials. The error of the fit is reduced particularly for larger network size (Fig~\ref{fig:bethe_large}\textbf{B} dashed lines), suggesting that the limited amount of data is mainly responsible for the estimation error. 
\par
In general, the estimation error is largest at time points where the parameters $\boldsymbol{\theta}_t$ change rapidly. This is a general problem of smoothing algorithms, including spike rate estimation, which depend on fixed smoothness parameter(s) (i.e., here $\lambda$) optimized for an entire observation period (see e.g., \cite{shimazaki2010kernel} for optimizing a variable smoothness parameter to cope with such abrupt changes).  

\bigskip \noindent 
\textbf{Comparison between Bethe and TAP approximation} To this end, only the Bethe approximation was used in combination with the pseudolikelihood to fit the model approximately. However, as discussed previously, the TAP approximation constitutes a potential alternative. To assess the quality of both approximations, we investigated a small network (15 neurons, 500 time bins, 1000 trials). The data was generated as described for Fig~\ref{fig:bethe_example}. The smaller network is considered because it allows to fit the model by an exact method without the Bethe or TAP approximations. Here the exact method refers to the method in which the expectation parameters are calculated exactly at the gradient search for the MAP estimates of model parameters (Eq \ref{eq:dlog_posterior}). It should be noted that we approximate the posterior density by the Gaussian distribution even for the ``exact method'' in the recursive Bayesian algorithm. Comparison of the approximation methods with the exact method determines the error that is caused by the approximation methods and not by the finite amount of data.
\par
\begin{figure}[tb]
\centering
\includegraphics[width=1\textwidth]{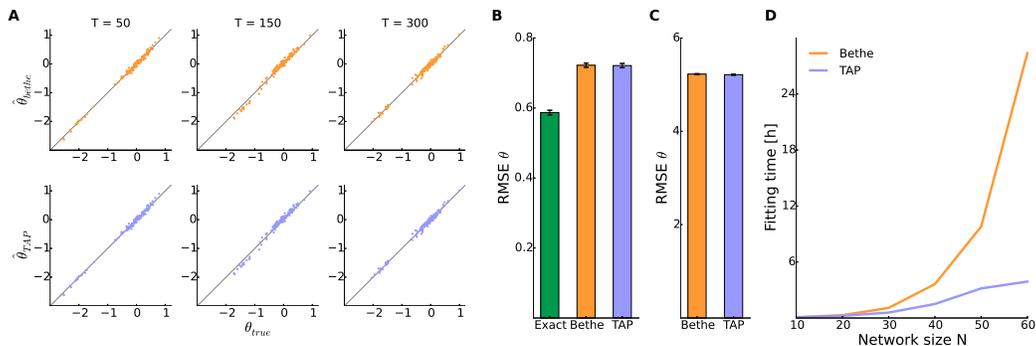}
\caption{ {\bf Comparison of the Bethe and TAP approximation.} Simulated neural activity composed of $500$ time bins, and $1000$ trials are used to compare the two approximation methods. The underlying model parameters follow Fig~\ref{fig:bethe_example}.
\textbf{A} Top: Ground truth $\boldsymbol\theta_t$ of a network of $15$ neurons vs. its smoothed estimate by pseudolikelihood-Bethe approximation at three different time points ($t=50,150,300$). Bottom: The same as above obtained with pseudolikelihood-TAP approximation. \textbf{B} The RMSE between the true model parameter $\boldsymbol\theta_t$ and its smoothed estimate by the exact inference, pseudolikelihood-Bethe, or pseudolikelihood-TAP approximation. The bar height and error bars indicate the mean and standard deviation from 10 realizations of data, each sampled from the same underlying parameters (generated as in Fig~\ref{fig:bethe_example}). \textbf{C} As in B the RMSE of the estimated model parameters for a network of $60$ neurons, composed of 6 equally sized subnetworks. \textbf{D} Running time as function of network size for the two different approximation methods.}\label{fig:tap_theta}
\end{figure}

First, investigation of three exemplary time points (Fig~\ref{fig:tap_theta}\textbf{A}) reveals that both the pseudolikelihood-Bethe and the pseudolikelihood-TAP approximation recover the underlying parameters. We examine the error across time bins by the RMSE. Comparing RMSE of the approximation results with the exact fit (Fig~\ref{fig:tap_theta}\textbf{B}) demonstrates that the both approximations perform worse in the same range.  To examine the approximations also for large networks ($N=60$) we sampled $1000$ trials (as for Fig~\ref{fig:bethe_large}). In Fig~\ref{fig:tap_theta}\textbf{C} we observe that errors of the approximations are comparable. Furthermore, we compare running times required for fitting the network of the two methods (Fig~\ref{fig:tap_theta}\textbf{D}). The pseudolikelihood-TAP approximation turns out to be faster than Bethe. We observed that the EM algorithm required more iterations for the Bethe approximation. Furthermore, the occasional use of the CCCP contributed to the long fitting time of the pseudolikelihood-Bethe procedure. 

\par
\begin{figure}[tb]
\centering
\includegraphics[width=1\textwidth]{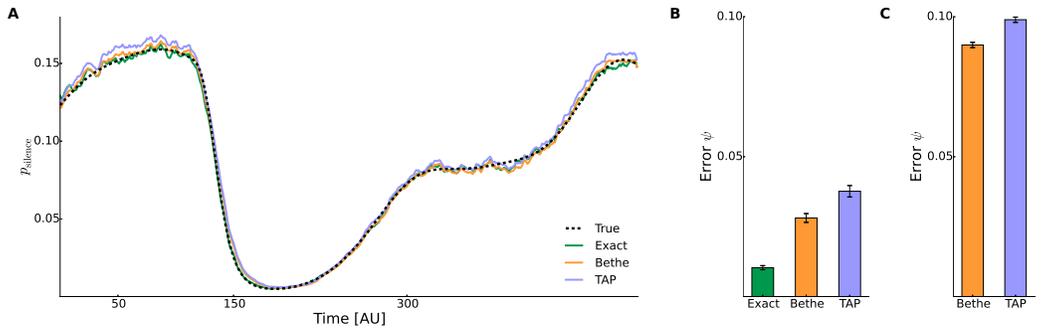}
\caption{ {\bf Time-varying probability of simultaneous silence.} Results of different approximation methods. The underlying model parameters are the same as in Fig~\ref{fig:tap_theta}. \textbf{A} The probability of simultaneous silence ($p_{\rm{silence}}(t) = \exp(-\psi_t)$) for a network of $15$ neurons as a function of time. The pseudolikelihood-Bethe (orange) and pseudolikelihood-TAP (lavender) method estimate the underlying value with sufficient accuracy (dashed black). For comparison, an estimate by the exact method (green) is shown. \textbf{B} The error between the approximate and true free energy $\psi_t$. \textbf{C} The error of free energy $\psi_t$ for large networks ($N=60$, data same as in Fig~\ref{fig:tap_theta}\textbf{C}).}\label{fig:tap_psi}
\end{figure}

Since both, Bethe and TAP, provide an approximation for the log partition function $\psi_t$ (Eq~\ref{eq:psi_TAPapprox_main} and~\ref{eq:bethe_psi_main}), we assess their performance for the same data as in Fig~\ref{fig:tap_theta}. The time evolution of simultaneous silence (directly linked to $\psi$ by Eq~\ref{eq:probsilence}) is recovered by exact, Bethe, and TAP (Fig~\ref{fig:tap_psi}\textbf{A}). The results show that the TAP approximation slightly overestimated the probability in this example. This is also reflected in the $\rm{Error}[\psi(\lbrace\hat{\boldsymbol\theta}_{t\vert T}\rbrace_t)]$ (Fig~\ref{fig:tap_psi}\textbf{B}), where the Bethe approximation performs better than the TAP method. However, the error for the Bethe approximation increases compared to the exact method. The relation between the two approximation methods persists also for large networks (Fig~\ref{fig:tap_psi}\textbf{C}). Another disadvantage of the TAP approximation is that the system of non-linear equations occasionally could not be solved. This happens more frequently when fitting larger networks and/or networks with stronger interactions. 
Therefore, it seems that the pseudolikelihood-Bethe approximation exhibits more accurate estimates; hence we will use it again for the following analysis. However the faster fitting of pesudolikelihood-TAP can be advantageous elsewhere.

\begin{figure}[t]
\begin{center}
\includegraphics[width=\textwidth]{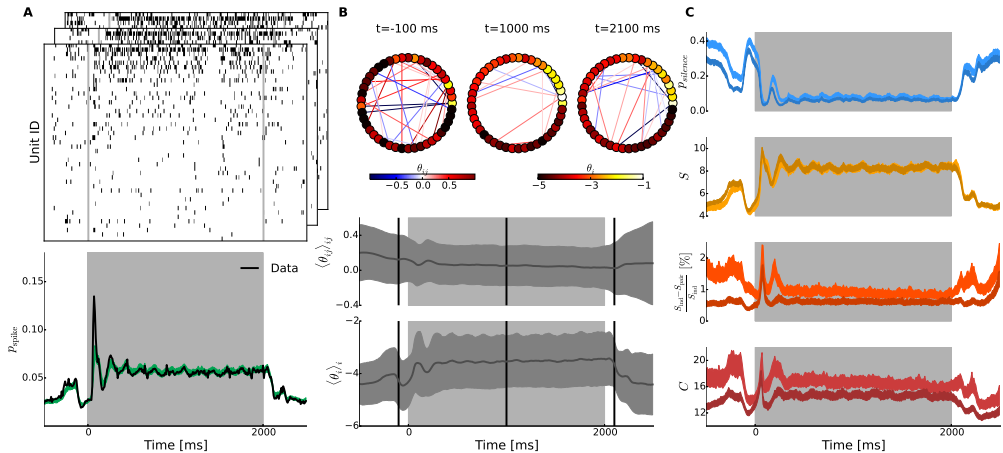}
\end{center}
\caption{{\bf Dynamic network inference from monkey V4 data.} In this experiment, a $90^\circ$ grating on a screen was presented to the monkey for $2 \mathrm{s}$ (light gray shaded areas). $1004$ trials were recorded, and binary spike trains were constructed with bin width of $10\ \mathrm{ms}$. \textbf{A} Top: Exemplary spiking data ($N=45$). Bottom: Empirical probability (black) of observing a spike over time and spike probability of the fitted model (green). \textbf{B} Top: The fitted network at three different time points, before, during, and after stimulation. Edges with significantly non-zero $\theta_{ij}^t$ are displayed (as in Fig~\ref{fig:bethe_example}). Bottom: The mean of smoothed MAP estimates for $\theta_{i}^t$ and $\theta_{ij}^t$ (dark gray line). The shaded area is the mean $\pm$ standard deviation. \textbf{C} Credible intervals of macroscopic measures of the network over time obtained from the smoothed estimates of the model (light color). Dark shaded area corresponds to the credible intervals of the estimates for trial shuffled data.}\label{fig:data1}
\end{figure}

\bigskip \noindent
\textbf{Dynamic network inference from V4 spiking data of behaving monkey} We now apply the approximate inference method to analyze activity of monkey V4 neurons recorded while the animal performed repeatedly ($1004$ trials) the following behavioral task. Each trial began when the monkey fixated its gaze within 1 degree of a centrally-positioned dot on a computer screen. After 150 ms, a drifting sinusoidal grating was presented for 2 s in the receptive field area of the neuronal population that was recorded, at which time the grating stimulus disappeared and the fixation point moved to a new, randomly chosen location on the screen, and the animal made an eye movement to fixate on the new location. Data epochs from 500 ms prior to grating stimulus onset until 500 ms after stimulus offset were extracted from the continuous recording for analysis. The spiking data obtained by micro-electrode recordings includes $112$ single and multi units identified by their distinct wave forms. The experiment was performed at the University of Pittsburgh. All experimental procedures were approved by the University of Pittsburgh Institutional Animal Care and Use Committee, and were performed in accordance with the United States' National Institutes of Health (NIH) \emph{Guide for the Care and Use of Laboratory Animals}. For details on experimental setup, recording and unit identification see \cite{snyder2015global}. The recorded units are tested for across-trial stationarity (which is the assumption of the model): The mean firing rates for each trial are standardized and if more than $5\%$ of the trials were outside the $95\%$ confidence interval the unit is excluded. After this preprocessing $45$ units remained. To obtain the binary data, the spike trains are discritized into time bins with $\Delta = 10\ \rm{ms}$ resulting into $300$ time bins over the course of the trial. Exemplary data are displayed in Fig~\ref{fig:data1}\textbf{A} Top. We note that the following conclusions of this analysis do not change even if we use smaller and larger bin size ($\Delta = 5\ \rm{and}\ 20\ \rm{ms}$). 

\par 
After the data are preprocessed, we analyze the network dynamics of the $45$ units during the task period by the state-space model for the neural population activity. Inference is done by using the pseudolikelihood-Bethe approximation. The results of fitting the state-space model are displayed in Fig~\ref{fig:data1}\textbf{B}. Before presenting detailed results, we note that considering dynamics in activity rates and neural correlations better explains the population activity while avoiding overfitting, compared to assuming that they are stationary. To assess this, we compared the predictive ability of the state-space model with that of the stationary model, using the Aikake (Bayesian) Information Criterion (AIC) \cite{akaike1974new} defined as $-2 l(\mathbf{X}^{1:T} |\mathbf{w}) + 2 k$, where $k$ is the number of free parameters in $\mathbf{w}$. To obtain the latter, we fitted the state-space model once more but now fixing $\lambda^{-1}=0$, which results in a stationary model since the state model in Eq~\ref{eq:statemodel} no longer contains variability. The result confirms that the dynamic model better predicts the data ($\rm{AIC}_{\rm{dyn}}=4467026$ for the dynamic model and $\rm{AIC}_{\rm{stat}}=4576544$ for the stationary model).
\par 
We observe stimulus locked oscillations in the population firing rate that are also captured by the model (Fig~\ref{fig:data1}\textbf{A} Bottom). The average of the estimated natural parameters (Fig~\ref{fig:data1}\textbf{B} Bottom) show that these oscillations are explained by the first order parameters $\theta_i^{t|T}$. We note that these oscillations are mainly caused by two units with high firing rates and they should not be considered as a homogeneous property of the network. Investigation of the network states before, during, and after the stimulus (Fig~\ref{fig:data1}\textbf{B} Top) reveals that the interactions $\theta_{ij}^{t|T}$ are altered over time. This is also reflected in an average over the all pairwise interactions (Fig~\ref{fig:data1}\textbf{B} Center), where the mean decreases during the stimulus presentation as well as the standard deviation. Thus neurons are likely to decorrelate during the stimulus presentation whereas the population rate increases and oscillates at the same time.
\par
Similarly to the analysis of artificial data (Fig~\ref{fig:bethe_example}), we measure the macroscopic properties of the fitted model over the task period (see Fig~\ref{fig:data1}\textbf{C} for credible intervals). To test the contribution of interactions in the recorded data, the model is once again fitted to trial shuffled data \cite{grun2009data}, which should destroy all correlations among units that do not occur due to chance. Comparison of the macroscopic measures between the models fitted to the original data and to the trial shuffled data shows how interactions among units alter the results. In the following, we will refer to the two models as ``actual'' and ``trial shuffled'' model.
\par 
The probability of simultaneous silence shows again the stimulus locked oscillations, and decreases during the stimulus period. The difference between the actual and trial shuffled model before the stimulus is larger than during and after the stimulus, suggesting that the observed positive interactions contributed to increasing the silence probability in particular before and after the stimulus period. The entropy reflects the oscillations and shows a strong increase ($\sim 1/3$) during the stimulus period. This is reasonable because we observe an increase in activity rates and a decrease in correlations - both effects should result in an increase in entropy. Next, we examine how much of the entropy is explained by the interactions among the neurons. To do so, at each time point we calculate the corresponding independent model by projecting the fitted interaction model to the independent model (i.e., the model with the same individual firing rates $\eta_i^t$ but with all $\theta_{ij}^t=0$). The entropy of the independent model $S_{\rm{ind}}$ should always be larger than $S_{\rm{pair}}$, the entropy of the model with interactions. Hence, a fraction of entropy explained by the interactions can be calculated as
\begin{equation}\label{eq:explained_entropy}
\frac{S_{\rm{ind}}-S_{\rm{pair}}}{S_{\rm{ind}}}.
\end{equation} 
In general, contribution of interactions to the entropy is small for these data ($\leq 2\%$). However, the contribution is less during stimulus presentation, compared to the period before the stimulus. Only in the beginning of the stimulus presentation, two peaks of correlated activity can be observed. The observed reduction of the fractional entropy for interactions could be caused by the increase of the first order parameters $\theta_i^{t}$ and/or by the decrease of the interactions $\theta_{ij}^{t}$ during the stimulus period. The decorrelation observed during the stimulus period is successfully dissociated from the oscillatory activity: Previously observed oscillations are absent in this measure of interactions. This result is important because ignoring such firing rate dynamics often leads to erroneous detection of positive correlations among neurons. A clear exception is the first peak appeared during the stimulus presentation, which was also observed in the trial-shuffled model. Indeed, the first sharp increase of the spike rates was not faithfully captured by the models, which caused spurious interactions in the trial-shuffled model. Last, the sensitivity (heat capacity) of the network over time is obtained. While for the artificial data in Fig~\ref{fig:bethe_example} the sensitivity showed a drastic decrease, such reduction is not observed in the V4 data. The sensitivity of the network is maintained at approximately the same value before and during the stimulus period. This is interesting since we already observed that before and during the stimulus the network seems to be in two qualitatively different states (low vs. high firing rate and strong vs. weak interactions).  After stimulus presentation the sensitivity drops. Overall, neural interactions contribute to have higher sensitivity (see light vs. dark credible intervals).

\begin{figure}[t]
\begin{center}
\includegraphics[width=\textwidth]{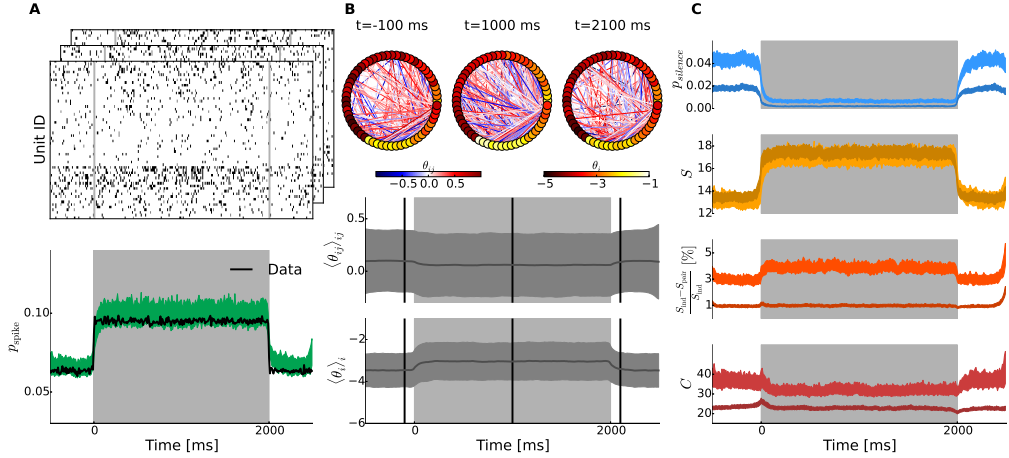}
\end{center}
\caption{{\bf Dynamic network inference from simulated balanced network data.} $60$ neurons ($40$ excitatory, $20$ inhibitory) are recorded from a simulated balanced network of 1000 leaky integrate-and-fire neurons that receive inputs from $800$ excitatory orientation selective Poisson neurons (mean firing rate $7.5\ \mathrm{Hz}$ when no stimulus present). See main text for the details. Stimulus was presented for $2\ \mathrm{s}$, and $1000$ trials are generated. Bin width is $10\ \mathrm{ms}$. The structure of this figure is the same as in Fig~\ref{fig:data1}.}\label{fig:balanced1}
\end{figure}

\bigskip \noindent
\textbf{Dynamic network inference from simulated balanced network data} Networks with balanced excitation and inhibition have been used to describe cortical activity\cite{amit1997model,van1996chaos}. To see whether the balanced network model can reproduce the findings from the recorded V4, we simulate spiking data using the balanced spiking network following \cite{renart2010asynchronous}, and analyze these data with the state-space model. The network consists of $1000$ leaky integrate-and-fire neurons ($800$ excitatory, $200$ inhibitory) (For details see \nameref{SI:text 4}). Connection probability is $20 \%$, between all neurons. The network receives input from $800$ Poisson neurons. Each input neuron has a Gaussian tuning curve, where the preferred direction is randomly assigned. We choose an experimental paradigm which resembles one of the V4 data. $1000$ trials of  $3\ \mathrm{s}$ duration are simulated. Before each trial, the simulation runs for $500\ \mathrm{ms}$ under random Poisson inputs such that the network state at the beginning of each trial is independent. Then the trial starts at $-500\ \mathrm{ms}$. At $0\ \mathrm{ms}$ a $90^\circ$ is shown for $2\ \mathrm{s}$ followed again by a $500\ \rm{ms}$ period of stimulus absence. The activity of $140$ neurons are recorded for investigation. From the recorded subpopulation, we further selected $40$ excitatory and $20$ inhibitory neurons with the highest firing rates for the following analysis. Binary spike trains were obtained by binning with $\Delta=10\ \mathrm{ms}$. Exemplary data are shown in Fig~\ref{fig:balanced1}\textbf{A} (top spike trains are from excitatory, and bottom spike trains from inhibitory neurons). We then fitted the state-space model to these data. 
\par

As for the V4 data, we show in Fig~\ref{fig:balanced1}\textbf{B} $3$ snapshots of the network ($N=60$) (Top), as well as mean and standard deviation of $\theta_i^{t \vert T}$ and $\theta_{ij}^{t \vert T}$ (Bottom). In contrast to the V4 network there are numerous significant non-zero couplings. However, similarly to the monkey data, we observe an increase for $\theta_i^t$ and a decrease of $\theta_{ij}^t$ during the stimulus period. We also assess the macroscopic states for the balanced network (Fig~\ref{fig:balanced1}\textbf{C}). As in the V4 data the probability of silence decreases during the stimulus period. Furthermore, compared to the trial shuffled result, the difference is larger before and after the stimulus than during the stimulus, suggesting a larger contribution of the couplings to silence when no stimulus is present. The entropy increases during the stimulus period. The credible interval for the trial shuffled data is narrower than for actual model and the entropy tends to be larger. Up to this point we did not find, in the macroscopic properties, significant qualitative differences between the V4 data and the simulated data from the balanced network. However, the entropy that is explained by the couplings increases during the stimulus, while in the V4 data a decrease is observed (Fig~\ref{fig:balanced1}\textbf{C}, third panel). Hence, the interactions in the balanced network become stronger during the stimulus, even though the mean of the couplings $\theta_{ij}^{t \vert T}$ decreases for this period. This can be explained by more negative values in estimated couplings during the stimulus period. The sensitivity slightly decreases when the stimulus is shown and, as for the V4 data, couplings contribute to higher sensitivity.
\par
\begin{figure}[t]
\begin{center}
\includegraphics[width=\textwidth]{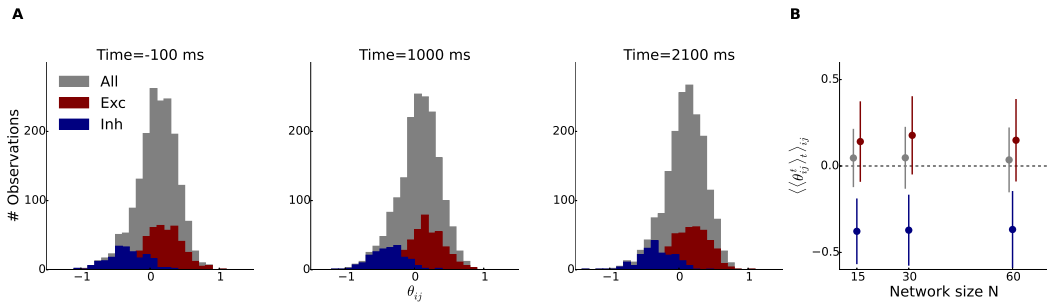}
\end{center}
\caption{{\bf Comparison of model interactions with synapses in the balanced network.} The synaptic structure is reflected in the inferred interactions. \textbf{A} Histograms of the interactions $\theta_{ij}^{t|T}$ for all pairs (gray), pairs that are connected by \textit{at least} one excitatory synapse (red), and those that are connected by \textit{at least} one inhibitory synapse (blue) at three different time points. \textbf{B} Averages of the couplings $\theta_{ij}^{t|T}$ across time and pairs as a function of a network size (always consisting of two thirds of excitatory and one third of inhibitory neurons). Colors as in A, and error bars denote standard deviations.}\label{fig:balanced2}
\end{figure}

 
Observing the dynamics in the model parameters poses the question how the actual synaptic connectivity structure of the network is reflected in the inferred interactions. Do positive values correspond to excitatory synapses, and negative to inhibitory ones? While for the V4 data this is impossible to assess, we compare the values of $\theta_{ij}^{t \vert T}$ of pairs, that are at least connected by one excitatory synapse and those that are connected by at least one inhibitory synapse (Fig~\ref{fig:balanced2}\textbf{A}, red and blue histograms respectively). In general, excitatory connected pairs show more positive values, while inhibiting ones tend to be negative. The most negative values are almost exclusively explained by inhibiting pairs. However, compared to all $\theta_{ij}^{t \vert T}$ (gray histogram) many positive couplings $\theta_{ij}^{t \vert T}$ do not represent excitatory connected pairs. Thus it is difficult to identify excitatory synapses from the inferred couplings. The result that inhibitory pairs showed stronger negative couplings, while excitatory pairs were mostly represented by weak positive couplings, can be explained by on average much stronger conductance of inhibitory synapses. 

\par
Finally we compare the mean values of couplings between different network sizes (Fig~\ref{fig:balanced2}\textbf{B}). To do so networks of size $N=15,30,60$ are fitted, where the network always consisted of one third inhibitory and two thirds excitatory neurons. However, neither for excitatory, inhibitory or all couplings we could identify dependency on the network sizes that can be analyzed by our model.
\par 

\section*{Discussion}
This study provides approximate inference methods for simultaneously estimating neural interactions of a large number of neurons, and quantifying macroscopic properties of the network in a time-resolved manner. We assessed performance of these methods by using simulated parallel spike sequences, and demonstrated the utility of the proposed approach by revealing dynamic decorrelation of V4 neurons and maintained susceptibility during stimulus presentations. Furthermore we compared those findings with data from a simple balanced network of LIF neurons, which suggested that further refinements were necessary to reproduce the observed network activity.

% Rate and correlation dynamics
Accurate assessment of correlated population activity in ongoing and evoked activity is a key to understand the underlying biological mechanisms and their coding principles. It is critical to model time-dependent firing rates to correctly assess neural interactions. If we apply a stationary model of neural interactions to independent neurons with varying firing rates, we may erroneously observe excess of correlations \cite{brody1999correlations,grun2009data,renart2010asynchronous,tyrcha2013effect,mochol2015stochastic}. Such an apparent issue of a stationary model can introduce considerable confusion in search of fundamental coding principles of neurons. Several related studies accounted for the nonstationary activity by modeling time-dependent external fields (c.f., $\{\theta_i^t\}$ in Eq~\ref{eq:observationmodel}) while fixing pairwise interactions  \cite{granot2013stimulus,tyrcha2013effect}. 
In addition to the external fields, however, we consider that modeling dynamics of correlations are important particularly for analyses of neurons recorded from awake animals because neural correlations are known to appear dynamically in relation to behavioral demand to the animals  \cite{vaadia1995dynamics,riehle1997spike,steinmetz2000attention,sakurai2006dynamic,shimazaki2012state}. 
Indeed, we found dynamic decorrelation of V4 neurons during stimulus presentation (Fig~\ref{fig:data1}C 3rd panel), which may reflect asynchronous neural activities under stimulus processing of an alert animal \cite{poulet2008internal,tan2014sensory}. In general, it is important to compare the result with that of surrogate data in which one destroys correlations to examine potentially short-lasting time-varying interactions in relation to behavioral paradigms.

% Relations to kinetic Ising and GLM
The current state-space model presumes that the neural dynamic follows a \textit{quasistatic} process. At each time $t$, we assumed that population activity is sampled from the \textit{equilibrium} joint distribution given by Eq~\ref{eq:observationmodel} across trials while the state of population activity smoothly changes within a trial. This is of course a simplified view of neuronal dynamics. Most notably, dependency of the neurons’ activity on their past activity makes the system a nonequilibrium one. Such activity is captured by models via the history effect, e.g., using the kinetic Ising model \cite{roudi2011mean,zeng2013maximum,tyrcha2013effect,dunn2013learning} or generalized linear models (GLM) of point and Bernoulli processes \cite{brillinger1988maximum,chornoboy1988maximum,pillow2008spatio,truccolo2005point}. Given the past activities, these models construct the joint activity assuming their conditional independence. The equilibrium and non-equilibrium models thus assume different generative processes, even though the pseudo-likelihood approximation for our equilibrium Ising model used similar conditional independence given the activity of other neurons at the same time. It is an important topic to include both modeling frameworks in the sequential Bayes estimation to better account for dynamic and nonequilibrium properties of neural activity \cite{shimazaki2013single}. The model goodness-of-fit may be additionally improved by including sparseness constraints on the couplings as was done in the stationary models \cite{stevenson2009bayesian,kolar2010estimating,koster2014modeling}. 

% About approximation methods to increase the number of neurons
In this study, we employed the classical pseudolikelihood method to perform MAP estimation of interactions (i.e., natural parameters) without computing the partition function. For the inverse problem without the prior, we may use alternative approximation methods such as Bethe and TAP approximations, and further state-of-the-art methods such as the Sessak-Monasson \cite{sessak2009small}, minimum-probability-flow \cite{sohl2011new}, and adaptive-cluster expansion \cite{cocco2012adaptive} method. However, here we chose the pseudolikelihood method because it was not trivial to apply the other methods to the Bayesian estimation. Alternatively, the Bethe and TAP approximation methods may be used to approximate the expectation parameters during the iterative procedure of the exact MAP estimation (Eq~\ref{eq:dlog_posterior}) because these methods allow us to estimate the expectation parameter from the natural parameters (the forward problem). However, as we found in the estimation of the Fisher information, TAP may occasionally fail and Bethe approximation by BP may not converge. Thus we rather used these methods after the MAP estimation was found by the pseudolikelihood method. The framework, however, is not limited to these approximation methods, and new methods may be incorporated into the state-space model to further increase the number of neurons that can be analyzed. 

It should be noted that the current model does not include higher-order interactions to explain the population dynamics. While neural higher-order interactions are ubiquitously observed \textit{in vivo} \cite{montani2009impact,ohiorhenuan2010sparse,yu2011higher,shimazaki2012state} as well as \textit{in vitro} \cite{ganmor2011sparse,tkavcik2013simplest,tkavcik2014searching,shimazaki2015simultaneous} conditions, it remains to be elucidated how they contribute to characterizing evoked activities. It is an important step to include higher-order interactions in the large-scale time-dependent model. However, the proposed method that includes up to pairwise interactions can be used as a null model for testing activity features involving higher-order interactions. For example, both experimental and modeling studies showed that simultaneous silence of neurons constitutes a major feature of higher-order interactions of stationary neural activities \cite{macke2009generating,shimazaki2015simultaneous}. It remains to be tested, though, if silence probability of all neurons recorded from behaving animals exceed prediction by the pairwise model. Such sparse population activity may be expected when animals process natural scenes, compared to artificial stimuli \cite{froudarakis2014population}.

% Conclusion
The limiting factor for the current model on the network size is rather the lack of data than the performance of the approximation methods (Fig~\ref{fig:bethe_large}). Hence, the state-space or other time-resolved methods that include dimension reduction techniques will be important approaches to explain activity of much larger populations than analyzed here. While there is still room for improvement, the currently proposed method already allows researchers to start testing hypotheses of network responses under distinct task conditions or brain states. These observations will serve to construct biophysical models of neural networks by constraining them, therefore revealing their coding principles.

\paragraph*{S1 Text}\textbf{Bethe approximation.}
\label{SI:text 1}
The Bethe approximation, belief propagation (BP), and concave convex procedure (CCCP) are well explained by \cite{yedidia2001idiosyncratic,yedidia2003understanding,yuille2002cccp}. However, for the sake of consistency the methods are summarized here once more. First the Bethe approximation in general will be discussed and subsequently the two algorithms to find its solution.
\par
The Bethe approximation is a variational approach. One assumes that the joint distribution of the Markov network can be written in terms of its individual and pairwise marginals
\begin{eqnarray}\label{eq:bethejoint}
q(\mathbf{x}) = \frac{\prod_{i,j>i}q(x_i,x_j)}{\prod_i q(x_i)^{N_i-1}},
\end{eqnarray}
where $N_i$ is the number of neighbors of neuron $i$. Eq~\ref{eq:bethejoint} ignores any cycles in the network and would be exact for a tree. The aim is to find the distribution $q(\mathbf{x})$ that is closest to our actual one $p(\mathbf{x})=\exp(\boldsymbol{\theta}^\prime\mathbf{F}(\mathbf{x}) - \psi)$, i.e., the one that minimizes the Kullback-Leibler (KL) divergence
\begin{equation}\label{eq:bethediv}
D_{\rm KL}(q\Vert p) = \sum_{\mathbf{x}}q(\mathbf{x})\log \frac{q(\mathbf{x})}{p(\mathbf{x})} = \phi(q) - \boldsymbol\theta^\prime\left\langle \mathbf{F}(\mathbf{x})\right\rangle_q + \psi,
\end{equation}
where $\langle \centerdot \rangle_q$ is the expectation over $q(\mathbf{x})$ and $\phi(q)$ its negative entropy. The Bethe approximation of the log partition function is given by 
\begin{eqnarray}\label{eq:bethepsi}
\psi \approx \psi_{\textrm{Bethe}} = \psi - D_{\rm KL}(q\Vert p) = -\phi(q) + \boldsymbol\theta^\prime\langle \mathbf{F}(\mathbf{x})\rangle_q.
\end{eqnarray}
Eq~\ref{eq:bethepsi} shows the nature of the approximation error. As long as the class of distribution $q(\mathbf{x})$ contains distributions close to the actual $p(\mathbf{x})$ the error will be small, because the KL divergence will be small. Furthermore, we see that $\psi_{\textrm{Bethe}}$ will underestimate $\psi$ systematically because $D_{\rm KL}\geq 0$. Eq~\ref{eq:bethediv} provides an objective function that needs to be minimized w.r.t. $q(\mathbf{x})$. Realizing that $\psi$ does not depend on $q(\mathbf{x})$, the problem is equivalent to maximizing Eq~\ref{eq:bethepsi}. Furthermore, $q(x_i)$ and $q(x_i,x_j)$ must fulfill following constraints:
\begin{equation}\label{eq:betheconstraints}
q(x_i) = \sum_{x_j}q(x_i,x_j) \mbox{ for } i={1,\ldots,N},j\neq i
\end{equation}
Normalization constraints for the marginals are ignored for the moment. The problem can be written as a Lagrangian
\begin{equation}\label{eq:bethelagrangian}
\mathcal{L}(q) = \psi_{\textrm{Bethe}} + \sum_{i\neq j}\sum_{x_i}\lambda_j(x_i)\left(\sum_{x_j}q(x_i,x_j) - q(x_i)\right).
\end{equation}
By setting the derivative w.r.t. $q(x_i)$ and $q(x_i,x_j)$ to $0$, the marginals can be expressed in terms of the Lagrangian multipliers
\begin{equation}\label{eq:bethemarginals}
\begin{split}
q(x_i, x_j) & \propto \exp\left(\theta_ix_i + \theta_jx_j + \theta_{ij}x_ix_j + \lambda_j(x_i) + \lambda_i(x_j)\right),\\
q(x_i) & \propto \exp\left(\theta_ix_i + \frac{\sum_{i\neq j}\lambda_j(x_i)}{N_i - 1}\right).
\end{split}
\end{equation}
This constitutes the Bethe approximation and it remains to find the marginals $q(x_i)$ and $q(x_i,x_j)$. In the following subsections two procedures are described, that diverge from this point.

\bigskip
\noindent
\textbf{Belief propagation} The BP starts from Eq~\ref{eq:bethemarginals}, but writes the Lagrangian multipliers in terms of messages as
\begin{equation}
\lambda_j(x_i) = \log \prod_{k\in N(i)\backslash j}m_{k}(x_i).
\end{equation}
$N(i)\backslash j$ are the set of neighbors of $i$ without $j$, and $m_{k}(x_i)$ is the \emph{message} sent from node $k$ to $i$. Substituting this into Eq~\ref{eq:bethemarginals} yields
\begin{equation}\label{eq:bethemarginalsmessages}
\begin{split}
q(x_i, x_j) & \propto \exp\left(\theta_ix_i + \theta_jx_j + \theta_{ij}x_ix_j\right)\prod_{k\in N(i)\backslash j}m_{k}(x_i)\prod_{k\in N(j)\backslash i}m_{k}(x_j),\\
q(x_i) & \propto \exp\left(\theta_ix_i\right)\prod_{k\in N(i)}m_{k}(x_i).
\end{split}
\end{equation}
By substituting these marginals into Eq~\ref{eq:betheconstraints} a set of self-consistent equations for the messages can be obtained
\begin{eqnarray}\label{eq:betheBP}
m_{j}(x_i) = \sum_{x_j}\exp\left(\theta_jx_j + \theta_{ij}x_ix_j\right)\prod_{k\in N(j)\backslash i}m_k(x_j).
\end{eqnarray}
The BP algorithm initializes the messages and solves Eq~\ref{eq:betheBP} iteratively until the algorithm converges. Having obtained the messages, the marginals can be computed by Eq~\ref{eq:bethemarginalsmessages} and they just need to be normalized in the end.

\bigskip
\noindent
\textbf{Concave convex procedure} While the BP algorithm takes care of the normalization constraints only in the end and hence does not sometimes converge, the CCCP \cite{yuille2002cccp} is more strictly about them, which guarantees convergence at the cost of computation time.
\par 
The starting point is the Lagrangian function depicted in Eq~\ref{eq:bethelagrangian}. Instead of maximizing $\psi_{\textrm{Bethe}}$ with the constraints, here we follow \cite{yuille2002cccp} that minimizes the Gibbs free energy, which is $-\psi_{\textrm{Bethe}}$. Furthermore, the normalization constraint
\begin{equation}\label{eq:CCCPnormalization}
\sum_{x_i,x_j}q(x_i,x_j) = 1,
\end{equation}
is added, resulting in the Lagrangian
\begin{align}\label{eq:CCCPlagrangian}
\mathcal{L}_{\textrm{CCCP}}(q) = 
  -\psi_{\textrm{Bethe}} + & \sum_{i\neq j}\sum_{x_i}\lambda_j(x_i)\left(\sum_{x_j}q(x_i,x_j) + q(x_i)\right) \nonumber\\
 & + \sum_{i\neq j}\gamma_{ij}\left(\sum_{x_i, x_j}q(x_i,x_j) - 1\right).
\end{align}
The basic principle of the CCCP is to realize that $-\psi_{\textrm{Bethe}}$ can be decomposed into a convex and a concave part
\begin{align}
-\psi_{\textrm{Bethe}} = & \underbrace{\sum_{i\neq j}\sum_{x_i,x_j} q(x_i,x_j)\log \frac{q(x_i,x_j)}{\exp (\theta_ix_i + \theta_jx_j + \theta_{ij}x_ix_j)}
+ \sum_i\sum_{x_i}q(x_i)\log \frac{q(x_i)}{\exp(\theta_ix_i)}}_{F_{\textrm{convex}}} \nonumber\\
& \underbrace{ - \sum_i N_i \sum_{x_i} q(x_i) \log \frac{q(x_i)}{\exp(\theta_ix_i)}}_{F_{\textrm{concave}}}.
\end{align}
Calculating the derivative w.r.t. the marginals yields the following iterative update rule for $q$
\begin{equation}\label{eq:CCCPderivates}
\begin{split}
\frac{\partial }{\partial q(x_i,x_j)}F_{\textrm{convex}}(q^{t+1}) =& -\frac{\partial }{\partial q(x_i,x_j)} F_{\textrm{concave}}(q^{t}) - \lambda_{i}(x_j) - \lambda_j(x_i) - \gamma_{ij}, \\
\frac{\partial }{\partial q(x_i)}F_{\textrm{convex}}(q^{t+1}) = &-\frac{\partial }{\partial q(x_i)} F_{\textrm{concave}}(q^{t}) + \sum_k \lambda_k(x_i).
\end{split}
\end{equation}
Note, that here $t$ is an integer describing the iterations of the algorithm and not the time-dependence of the model. By updating the marginals with Eq~\ref{eq:CCCPderivates}, $-\psi_{\textrm{Bethe}}$ monotonically decreases (see Theorem 2 in \cite{yuille2002cccp}). Writing the update explicitly for the marginals, we get
\begin{equation}\label{eq:CCCPmarginals}
\begin{split}
q^{t+1}(x_i,x_j) = & \exp\left(\theta_ix_i + \theta_jx_j +\theta_{ij}x_ix_j - \lambda_i(x_j) - \lambda_j(x_i) - \gamma_{ij}\right), \\
q^{t+1}(x_i) = & \left(\frac{q^t(x_i)}{\exp(\theta_ix_i)}\right)^{N_i}\exp\left(\theta_ix_i + N_i + \sum_j\lambda_j(x_i) - 1 \right).
\end{split}
\end{equation}
Assume we have a set of Lagrangian multipliers such that the constraints in Eq~\ref{eq:betheconstraints} and \ref{eq:CCCPnormalization} are satisfied. Then $-\psi_{\textrm{Bethe}}$ can be decreased by updating the marginals with Eq~\ref{eq:CCCPmarginals}. However, by doing so the constraints will be violated and one has to update the Lagrangian multipliers. By substituting Eq~\ref{eq:CCCPmarginals} into the constraints (Eq~\ref{eq:betheconstraints} and \ref{eq:CCCPnormalization}), one gets self-consistent equations for the multipliers that write as
\begin{equation}
\begin{split}
\exp(\gamma_{ij}) =& \sum_{x_i,x_j}\exp\left(\theta_ix_i + \theta_jx_j + \theta_{ij}x_ix_j - \lambda_j(x_i) - \lambda_i(x_j) - 1\right),\\
\exp\left(2\lambda_j(x_i)\right) =& \frac{\sum_{x_j} \exp(\theta_jx_j + \theta_{ij}x_ix_j - \lambda_i(x_j) - \gamma_{ij})}{\left(\frac{q^t(x_i)}{exp(\theta_ix_i)}\right)^{N_i}\exp\left(N_i + \sum_{k\neq\{i,j\}}\lambda_k(x_i)\right)}.
\end{split}
\end{equation}
The multipliers are updated sequentially until the constraints for the marginals are again satisfied. 
\par 
The CCCP always updates first the marginals. For each update the Lagrangian multipliers have to be updated until the constraints are fulfilled again. This alternating procedure is done until the Bethe free energy converges.

\paragraph*{S2 Text}\textbf{TAP approximation.}
\label{SI:text 2}
The \emph{Thouless-Anderson-Palmer} (TAP) approach is based on mean-field theory and was first suggested by \cite{thouless1977solution}. There are several ways to derive this approximation \cite{opper2001advanced}. Here we follow the lines of \cite{tanaka1998mean, tanaka1999theory} using the \emph{Plefka expansion} \cite{plefka1982convergence}. The major difference in our calculation is that $x\in\lbrace 0,1 \rbrace$ instead of $\lbrace -1,1 \rbrace$.
\par 
The Kullback-Leibler (KL) divergence between two probability mass functions is given by
\begin{equation}
D_{\rm KL}(q\Vert p) = \sum_{\mathbf{x}}q(\mathbf{x})\log\frac{q(\mathbf{x})}{p(\mathbf{x})},
\end{equation}
For the exponential family distribution $p(\boldsymbol{x})=\exp(\boldsymbol{\theta}_p^\prime\mathbf{F}(\mathbf{x}) - \psi_p)$, it is written as \cite{amari2001information} 
\begin{equation}
D_{\rm KL}(q\Vert p)
= \phi_q -\boldsymbol{\theta}_p^\prime \boldsymbol{\eta}_q +\psi_p,
\end{equation}
where $\phi_q$ is the negative entropy of $q(\mathbf{x})$ and $\boldsymbol{\eta}_q=\langle \mathbf{F}(\mathbf{x}) \rangle_{q}$. Here our goal is to find $p(\mathbf{x})$ that minimizes the KL divergence given $q(\mathbf{x})$. This is equivalent to maximizing $\boldsymbol{\theta}_p^\prime \boldsymbol{\eta}_q - \psi_p$. If $q(\mathbf{x})$ is an empirical distribution, this is also equivalent to maximizing likelihood of the model. Below we identify $\boldsymbol{\theta}_p$ with the one that maximizes the likelihood given $q(\mathbf{x})$. At this point, the expectation of $\mathbf{F}(\mathbf{x})$ by $p(\mathbf{x})$ is identical to $\boldsymbol{\eta}_q$ ($\langle \mathbf{F}(\mathbf{x}) \rangle_{p}=\langle \mathbf{F}(\mathbf{x}) \rangle_{q}$). Hence by dropping the subscripts, the maximized likelihood is written as  
\begin{equation}\label{eq:legendre1}
\phi(\boldsymbol\eta) = \boldsymbol\theta^\prime\boldsymbol{\eta}-\psi(\boldsymbol\theta),
\end{equation}
which is also the negative entropy of $p(\mathbf{x})$. We also note the relation:
\begin{equation}\label{eq:legendre2}
\frac{\partial \psi}{\partial \boldsymbol\theta}=\boldsymbol\eta.
\end{equation}
Eqs~\ref{eq:legendre1} and~\ref{eq:legendre2} represent the \emph{Legendre transform}: a translation of a functional relation from $\psi( \boldsymbol\theta)$ to $\phi(\boldsymbol\eta)$. 
\par
For our model $p(\mathbf{x})$ we now introduce a single scalar $\alpha$ into the distribution which controls the strength of interactions
\begin{equation}\label{eq:p_alpha}
p(\mathbf{x}) = \exp\left(\sum_i\theta_ix_i + \frac{\alpha}{2}\sum_{i\neq j}\theta_{ij}x_ix_j - \psi\right).
\end{equation}
The model becomes an independent model when $\alpha=0$. Here the log partition function is a function of $\{\theta_{i}\}$ and $\{\alpha \theta_{ij}\}$. We now change the variables $\{\theta_i\}$ to $\{\eta_i\}$ by the Legendre transformation of the log partition function to obtain a new free energy:
\begin{equation}
\tilde{\phi}(\{\eta_i\},\{\alpha \theta_{ij}\}) = \sum_i{\theta_i \eta_i} - \psi(\{\theta_i\},\{\alpha \theta_{ij}\}).
\end{equation}
The function $\tilde{\phi}$ is a function of $\eta_i$, $\theta_{ij}$, and $\alpha$. By assuming weak pairwise interactions because of small $\alpha$, we approximate $\tilde{\phi}$ by expanding it around the independent model:
\begin{equation}
\tilde{\phi}(\alpha) = \tilde{\phi}\vert_{\alpha=0} + \left.\frac{\partial \tilde{\phi}}{\partial \alpha}\right\vert_{\alpha=0}\alpha + \frac{1}{2}\left.\frac{\partial^2 \tilde{\phi}}{\partial \alpha^2}\right\vert_{\alpha=0}\alpha^2 + \cdots
\end{equation}
The TAP approximation is obtained using expansions up to $\alpha^2$. By setting $\alpha=1$, the approximated free energy $\tilde{\phi}$ is obtained as
\begin{equation} \label{phi_TAPapprox}
\begin{split}
\tilde{\phi}(1) \approx & \sum_{i=1}^N \left(\eta_i \log \eta_i + (1-\eta_i)\log(1-\eta_i)\right) -\frac{1}{2}\sum_{j\neq i} \theta_{ij}\eta_i\eta_j \\
& -\frac{1}{8}\sum_{j\neq i} \theta_{ij}^2(\eta_i - \eta_i^2)(\eta_j - \eta_j^2).
\end{split}
\end{equation}
This approach is called the Plefka expansion method \cite{plefka1982convergence}. The first term is the negative entropy of the independent model whereas the second and third terms are obtained by computing derivatives of the negative entropy w.r.t. $\alpha$. Derivation of the last two terms are given as Eqs~\ref{eq:1deriv_alpha} and \ref{eq:2deriv_alpha} in the end of this section. 

By taking the derivative w.r.t. $\eta_i$ in Eq~\ref{phi_TAPapprox}, we obtain a system of self-consistent equations
\begin{equation}\label{eq:selfconsistentequations}
\theta_i = \log \left(\frac{\eta_i}{1-\eta_i}\right) - \sum_{j\neq i} \theta_{ij}\eta_j - \frac{1}{2}\sum_{j\neq i} \theta_{ij}^2\left(\frac{1}{2}-\eta_i\right)\left(\eta_j-\eta_j^2\right).
\end{equation}
Taking the derivative of Eq~\ref{eq:selfconsistentequations} w.r.t. $\eta_j$, we obtain the $(i,j)$ element of the inverse Fisher information matrix (for $\theta_{i}$s):
\begin{equation}\label{eq:tap_fisher}
[\mathbf{G}^{-1}]_{ij} =\frac{1}{\eta_i(1 - \eta_i)} \delta_{ij} - \theta_{ij} - \theta_{ij}^2\left(\frac{1}{2} - \eta_i\right)\left(\frac{1}{2} - \eta_j\right).
\end{equation}
Here $\delta_{ij}$ is the Kronecker delta function, where it is $1$ if $i=j$ and $0$ otherwise. We also let $\theta_{ii}=0$ for $i=1,\ldots,N$. Using these formulas we can solve the \emph{forward problem}, i.e., given $\boldsymbol\theta$ obtain an approximation for $\boldsymbol\eta$. First Eq~\ref{eq:selfconsistentequations} is solved numerically to get $\{\eta_i\}$. Then we obtain the upper left part of the inverse Fisher information matrix by Eq~\ref{eq:tap_fisher} and invert it. By Eq~16 (main text), we see that $\eta_{ij}$s are given by 
\begin{equation}
\eta_{ij}=[\mathbf{G}]_{ij}+\eta_i\eta_j.
\end{equation}

Finally, the inverse Legendre transformation yields the TAP approximation of the log partition function,
\begin{align} \label{psi_TAPapprox}
\psi_{\textrm{TAP}} \approx & \sum_i{\theta_i \eta_i} - \tilde{\phi}(1) \nonumber\\
=& \sum_i{\theta_i \eta_i} - \sum_{i=1}^N \left\{\eta_i \log \eta_i + (1-\eta_i)\log(1-\eta_i)\right\} +\frac{1}{2}\sum_{j\neq i} \theta_{ij}\eta_i\eta_j \nonumber\\
& +\frac{1}{8}\sum_{j\neq i} \theta_{ij}^2(\eta_i - \eta_i^2)(\eta_j - \eta_j^2).
\end{align}
Here we use $\{\eta_i\}$ obtained at Eq~\ref{eq:selfconsistentequations}. 

Below we compute derivatives of the negative entropy function. Let the hamiltonian of the system be $H = H^{ext} + \alpha H^{int}$, where $H^{ext}=-\sum_i\theta_ix_i$ and $H^{int}=-\frac{1}{2}\sum_{i\neq j}\theta_{ij}x_ix_j$. We reiterate that $\tilde{\phi}$ is a function of mixture coordinates $(\{\eta_{i}\},\{\alpha \theta_{ij}\})$ whereas $\{\theta_{i}\}$ and $H^{ext}$ are dependent on these parameters. 

The first derivative is given as
\begin{align}\label{eq:1deriv_alpha}
\frac{\partial \tilde{\phi}}{\partial \alpha} = & \sum_{i=1}^N \frac{\partial \theta_i}{\partial \alpha}  \eta_i - \frac{\partial}{\partial \alpha} \log\sum_{\mathbf{x}}\exp(-H) \nonumber\\
 = & \sum_{i=1}^N \frac{\partial \theta_i}{\partial \alpha}  \eta_i + \frac{1}{\sum_{\mathbf{x}}\exp(-H)}\sum_{\mathbf{x}}\exp(-H)\left[ H^{int} + \frac{\partial H^{ext}}{\partial \alpha}\right] \nonumber\\
 = & \sum_{i=1}^N \frac{\partial \theta_i}{\partial \alpha}  \eta_i + \sum_{\mathbf{x}}\exp(-H-\psi)\left[ H^{int} + \frac{\partial H^{ext}}{\partial \alpha}\right] \nonumber\\
 = &\sum_{i=1}^N \frac{\partial \theta_i}{\partial \alpha}  \eta_i + \left\langle H^{int}\right\rangle_\alpha + \left\langle \frac{\partial H^{ext}}{\partial \alpha}\right\rangle_\alpha  \nonumber\\
 = &\left\langle H^{int}\right\rangle_\alpha,
\end{align}
where $H = H^{ext} + \alpha H^{int}$ and $\langle\centerdot\rangle_\alpha$ is the expectation w.r.t. Eq~\ref{eq:p_alpha} which depends on $\alpha$. Substituting $\alpha=0$ yields
\begin{equation}\label{eq:1deriv_0}
\left. \frac{\partial \tilde{\phi}}{\partial \alpha}\right\vert_{\alpha=0} = \left\langle-\frac{1}{2}\sum_{j\neq i} \theta_{ij}x_ix_j\right\rangle_{\alpha=0} 
= -\frac{1}{2}\sum_{j\neq i} \theta_{ij}\eta_i\eta_j.
\end{equation}
The second derivative is given as
\begin{align}\label{eq:2deriv_alpha}
\frac{\partial^2 \tilde{\phi}}{\partial \alpha^2} = & \frac{\partial \tilde{\phi}}{\partial \alpha} \left\langle H^{int}\right\rangle_\alpha  = \frac{\partial}{\partial \alpha}\sum_{\mathbf{x}}\exp(-H-\psi) H^{int} \nonumber\\
 = & \sum_{\mathbf{x}}\left[\exp(-H-\psi)\left(- \frac{\partial}{\partial \alpha} H - \frac{\partial}{\partial \alpha} \psi\right) H^{int}\right] \nonumber\\
 = &  \left\langle\left(- \frac{\partial}{\partial \alpha} H - \left\langle -\frac{\partial H}{\partial \alpha}\right\rangle_\alpha \right) H^{int}\right\rangle_\alpha \nonumber\\
 = &  \left\langle \left(- \frac{\partial}{\partial \alpha} H^{ext} - H^{int} - \left\langle -\frac{\partial}{\partial \alpha} H^{ext} - H^{int}\right\rangle_\alpha \right) H^{int}\right\rangle_\alpha \nonumber\\
 = &  \left\langle \left( \sum_i \frac{\partial \theta_i}{\partial \alpha} x_i - H^{int} - \left\langle \sum_i \frac{\partial \theta_i}{\partial \alpha} x_i\right\rangle_\alpha + \left\langle H^{int}\right\rangle_\alpha\right) H^{int}\right\rangle_\alpha \nonumber\\
 = &  \left\langle\left(  \sum_i \frac{\partial \theta_i}{\partial \alpha}( x_i - \eta_i)- H^{int}  + \left\langle H^{int}\right\rangle_\alpha\right) H^{int}\right\rangle_\alpha.
\end{align}
Substituting $\alpha=0$ yields
\begin{align}\label{eq:2deriv_0}
\left. \frac{\partial^2 \tilde{\phi}}{\partial \alpha^2}\right\vert_{\alpha=0} = & \left\langle\left(  \sum_i \frac{\partial \theta_i}{\partial \alpha}( x_i - \eta_i)- H^{int}  + \left\langle H^{int}\right\rangle_{\alpha=0}\right) H^{int}\right\rangle_{\alpha=0} \nonumber\\
 = & \left\langle\left(  -\sum_{j\neq i} \theta_{ij}\eta_j( x_i - \eta_i)+ \frac{1}{2}\sum_{j\neq i} \theta_{ij}x_ix_j   - \frac{1}{2}\sum_{j\neq i}\theta_{ij}\eta_i\eta_j\right) H^{int}\right\rangle_{\alpha=0} \nonumber\\
 = & -\frac{1}{2}\left\langle\left(  -\sum_{j\neq i} \theta_{ij}\eta_jx_i + \frac{1}{2}\sum_{j\neq i} \theta_{ij}\eta_j\eta_i + \frac{1}{2}\sum_{j\neq i} \theta_{ij}x_ix_j  \right) \sum_{j\neq i} \theta_{ij}x_ix_j \right\rangle_{\alpha=0} \nonumber\\
 = & -\frac{1}{4}\sum_{j\neq i} \theta_{ij}^2(\eta_i - \eta_i^2)(\eta_j - \eta_j^2).
\end{align}
For the last equality we made use of:
\begin{align}\label{eq:deriv_theta_alpha}
\left. \frac{\partial \theta_k}{\partial \alpha}\right\vert_{\alpha=0} = & \left. \frac{\partial^2 \tilde{\phi}}{\partial \alpha \partial \eta_k}\right\vert_{\alpha=0} =  \left. \frac{\partial^2 \tilde{\phi}}{ \partial \eta_k\partial \alpha}\right\vert_{\alpha=0} \nonumber\\
= & \frac{\partial}{\partial \eta_k}\left\langle H^{int}\right\rangle_{\alpha=0} 
= -\frac{1}{2}\frac{\partial}{\partial \eta_k}\sum_{j\neq i} \theta_{ij}\eta_i\eta_j = -\sum_{j\neq k} \theta_{kj}\eta_j.
\end{align}

\paragraph*{S3 Text}\textbf{Generation of simulated data.} 
\label{SI:text 3}
Here we explain how the underlying model parameters for Figs 1-4 are generated, and how the artificial spike data is sampled from the model. We discuss the model parameters used to generate the subpopulation activity. Benefit of constructing a large network as combination of independent small subpopulations is that we can exactly compute macroscopic network states (sparsity, entropy, and heat capacity). An additional advantage is that one can exactly sample spiking data without utilizing Monte Carlo methods. Furthermore, this way we do not need to scale the standard deviation of interactions to compare different network sizes. 

In order to construct smooth dynamics, the underlying time-varying parameters $\boldsymbol\theta_{1:T}$ are sampled as Gaussian processes of $T=500$ time bins, for $i,j=1,\ldots,N$:
\begin{equation}
\begin{split}
\theta_i^{1:T} \sim \mathcal{GP}(\boldsymbol\mu, \mathbf{K}), \\
\theta_{ij}^{1:T} \sim \mathcal{GP}(\mathbf{0}, \mathbf{K}),
\end{split}
\end{equation}
where $\boldsymbol\mu$ is a mean vector of size $T$, and $\mathbf{K}$ is the $T \times T$ covariance matrix. For $\theta_i^{1:T}$, the mean vector $\boldsymbol\mu = (\mu_1,\ldots,\mu_T)$ is modulated using an inverse Gaussian function as 
\begin{equation}
\mu_t = 
\begin{cases}
-2 \mbox{ for } t<100\\
-2 + \frac{\lambda}{2\pi g(t)^3}\exp\left(- \frac{\lambda}{2f(t)}(g(t)-1)^2\right) \mbox{ for } t\geq 100,
\end{cases}
\end{equation}
where $g(t)=3(t - 100)/400$ and $\lambda=3$. For $\theta_{ij}^{1:T}$, the mean is fixed at zero. To produce smooth processes, the covariance matrix $\mathbf{K}$ dictating the smoothness for both $\theta_i$ and $\theta_{ij}$ is chosen as 
\begin{equation}
[\mathbf{K}]_{t,t^\prime} = \frac{1}{\sigma_1}\exp\left(-\frac{\vert t-t^\prime\vert^2}{2\sigma_2^2}\right),
\end{equation}
where $\sigma_1 = 12$ and $\sigma_2 = 50$. While the processes of the first order natural parameters $\{\theta_i^{1:T}\}$ have time-varying mean at different time points, it should be noted that the sampled interactions $\{\theta_{ij}^{1:T}\}$ also smoothly change over time. 

\paragraph*{S4 Text}\textbf{Simulated experiment with a balanced network.}
\label{SI:text 4}
To examine consequences of the proposed statistical analysis on a physiologically plausible model of cortical networks, we used a well-studied balanced network model (see Fig 3 of \cite{renart2010asynchronous}) with slight modifications. For the network simulations, we use the Brian simulator \cite{stimberg2014equation}. The network consists out of $3$ distinct populations: Input ($N=800$), excitatory ($N=800$) and inhibitory ($N=200$) neurons. Connectivity and parameters of the conductance-based leaky integrate-and-fire neurons are set as in the original work. In contrast to the cited paper the inputs provided to the network are inhomogeneous Poisson processes whose firing rates are all given by  
\begin{equation}
r(t) = 7.5\ \mathrm{Hz} + 5\cdot stim(t)\exp\left(-(\nu_{stim} - \nu_{pref})^2\right)\ \mathrm{Hz},
\end{equation}
where $stim(t)=1$ if a stimulus is present, and $0$ otherwise. $\nu_{stim}\in [-\pi,\pi)$ is the orientation of the stimulus. The preferred direction of each input is drawn from a uniform distribution $\nu_{pref} \sim [-\pi,\pi)$.
\par 
The following experiment was simulated $1000$ times with this network. A simulation started with activity without any stimulation for $1\ \mathrm{s}$. Then a stimulus with $\nu_{stim}=\pi$ is shown to the network for $2\ \mathrm{s}$. A period of $0.5\ \mathrm{s}$ in the absence of stimulus follows. 
\par
During the experiment the spike times of $100$ randomly selected excitatory and $40$ inhibitory neurons are recorded. For the statistical analysis the $40$ excitatory and the $20$ inhibitory neurons are chosen that exhibit the highest firing rates in the recorded population.

\section*{Acknowledgments} The authors thank Thomas Sharp for originally translating Matlab code written by HS to Python code, and Adam Snyder and Matthew A. Smith for kindly providing the V4 spiking data. CD and HS acknowledge Taro Toyoizumi for hosting CD's stay in RIKEN Brain Science Institute, and Timm Lochmann for valuable ideas and discussions.

\small
\bibliography{bibliography}

%\end{thebibliography}

\end{document}